\documentclass[draft]{IEEEtran}
\usepackage[utf8]{inputenc}
\usepackage{enumitem,kantlipsum}
\usepackage{fontawesome}
\usepackage{color}
\usepackage[]{footmisc}
\ifCLASSINFOpdf
   \usepackage[pdftex]{graphicx}
\else
\fi
\usepackage{amsmath,amsfonts,amsthm,bm,newunicodechar}

\newcommand{\RN}[1]{%
	\textup{\uppercase\expandafter{\romannumeral#1}}%
}

\newcommand\vcent[1]{\vcenter{\hbox{#1}}}
\newcommand\loudspeaker[1][3]{\ensuremath{\vcent{\rule{.6ex}{.6ex}}\kern-.5ex%
		\vcent{\scalebox{.6}[1]{\rotatebox[origin=center]{90}{$\blacktriangle$}}}%
		\ifnum#1>0\relax\kern.05ex\vcent{\scalebox{.4}{\ttfamily)}}%
		\ifnum#1>1\relax\kern-.4ex\vcent{\scalebox{.56}{\ttfamily)}}%
		\ifnum#1>2\relax\kern-.55ex\vcent{\scalebox{.7}{\ttfamily)}}%
		\fi\fi\fi}%
}
\DeclareSymbolFont{matha}{OML}{txmi}{m}{it}
\DeclareMathSymbol{\varv}{\mathord}{matha}{118}

\begin{document}
\title{A time-domain nearfield frequency-invariant beamforming method}
\author{Fei~Ma,~\IEEEmembership{Member,~IEEE,}
~Thushara D. Abhayapala,~\IEEEmembership{Senior Member,~IEEE},\\
and~Prasanga N. Samarasinghe,~\IEEEmembership{Senior Member,~IEEE}
\thanks{This work is sponsored by the Australian Research Council (ARC) Discovery Projects funding schemes with project 
numbers DP200100693.}
\thanks{Fei Ma, Thushara D. Abhayapala, and Prasanga N. Samarasinghe are all with the Research School of Engineering, 
College of Engineering and Computer Science, The Australian National University, Canberra, ACT 2601, Australia 
(e-mail: fei.ma, Thushara.Abhayapala, prasanga.samarasinghe@anu.edu.au).}
\thanks{Manuscript received xxx xxx, xxx; revised xxx xx, xxx.}}


\maketitle
\begin{abstract}
Most existing beamforming methods are frequency-domain methods, and are designed for enhancing a farfield target source over 
a narrow frequency band. They have found diverse applications and are still under active development. However, they struggle 
to achieve desired performance if the target source is in the nearfield with a broadband output. This paper proposes a 
time-domain nearfield frequency-invariant beamforming method. The time-domain implementation makes the beamformer output 
suitable for further use by real-time applications, the nearfield focusing enables the beamforming method to suppress an 
interference even if it is in the same direction as the target source, and the frequency-invariant beampattern makes the 
beamforming method suitable for enhancing the target source over a broad frequency band. These three features together make 
the beamforming method suitable for real-time broadband nearfield source enhancement, such as speech enhancement in room 
environments. The beamformer design process is separated from the sound field measurement process, and such that a designed 
beamformer applies to sensor arrays with various structures. The beamformer design process is further simplified by 
decomposing it into several independent parts. Simulation results confirm the performance of the proposed beamforming 
method. 
\end{abstract}

\IEEEpeerreviewmaketitle

\section{Introduction}
\IEEEPARstart{B}{eamforming} is a method that spatially filters the measurements of a number of sensors to 
form a response~\cite{Bvan1988}. 
Thanks to its capability to suppress the interference and preserve (or enhance) the target source~\cite{MSES}, 
beamforming has found applications in diverse areas, such as wireless communication~\cite{Bvan1988}, 
medical imaging~\cite{Bvan1997}, speech dereverberation~\cite{beamspeech} and recognition \cite{smartspeaker}. 

The hundreds of beamforming methods reported in literature~\cite{MSES} can be broadly 
classified into five categories: 
(1) frequency-domain or time-domain implementation;
(2) farfield or nearfield focusing;
(3) frequency-variant or frequency-invariant beampattern;
(4) robust or non-robust design;
(5) data-independent or data-dependent spatial filtering. 
Hereafter we introduce these five categories of beamforming methods.

The majority of existing beamforming methods implement the spatial filtering process in the frequency-domain. 
They use the discrete Fourier transform (DFT) (or specifically the fast Fourier transform (FFT)) to transform 
a frame of time-domain measurements into the time-frequency domain~\cite{DSP}, which allows better control 
of the beampattern within a narrow frequency band~\cite{MSES}. 
The time-domain methods, on the other hand, process the measurements in the time-domain directly and are 
inherently broadband~\cite{Yan2011}. 
Without resorting to the FFT, the time-domain filtering process can be computationally intensive. 
Nonetheless, the sample-by-sample filtering process makes the time-domain beamforming methods are of low 
latency (or zero latency).  

Depending on the location of the target source, a beamforming method may focus on the farfield or the nearfield. 
The wavefronts of a farfield source only differ by the relative phase. This fact simplifies the design of the spatial 
filters for a farfield beamforming method.
The wavefronts of a nearfield source, on the other hand, undergo both amplitude decay and phase shift. 
This fact makes it difficult to design the spatial filter for a nearfield beamforming method. 
Thus, most of the existing beamforming methods presume the farfield scenarios. Nonetheless, as one would expect, 
a farfield beamforming method not necessarily achieve the desired beampattern if the target source 
is in the nearfield~\cite{Thushara_near}. 

For a typically beamformer, the spectral response and spatial response are determined by a single set of filter 
coefficients, and thus are coupled with each other~\cite{coupling}. The coupling leads to a beampattern varying 
with the frequency, which is undesirable for a given interference and target source layout. A 
frequency-invariant beamformer, on the other hand, decouples the filter coefficients that are responsible for 
the spatial response from the filter coefficients that are responsible for the spectral response, and achieves 
a desired beampattern irrespective to the frequency~\cite{Dbward,Liu,Huang}.

In real world implementation of a beamforming method, there are always some imperfections, such as the 
imprecise knowledge of the target source location, sensor array design defects, and thermal noises~\cite{Rafaely2015}. 
These imperfections can impair a beamforming method, making it unable to achieve the desired performance. 
A robust design is to make the beamforming method invulnerable to such imperfections
at the expense of reduced performance such as the mainlobe width or the sidelobe level~\cite{rb1,rb2}. 

Depending on whether the statistics of the interference are taken into consideration, a beamforming method is classified 
as data-dependent or data-independent~\cite{MSES}. A data-dependent beamforming method, such as the minimum variance 
distortionless response, the linearly constrained minimum variance, and the multichannel Wiener filter, explicitly adapt 
the spatial filter according to the interference statistics, and tends to yield better performance 
\cite{MSES}. 
The data-independent beamforming method with its fixed spatial filter design tends to achieve inferior 
performance but is easier to implement~\cite{MSES}. 

We regard the time-domain implementation, nearfield focusing, frequency-invariant beampattern, 
robust design, and data-dependent spatial filters as desirable features. 
These features provide beamforming methods with additional capabilities or better performance, 
but can be at odds with each other. 
Thus, a specific beamforming method maybe designed to possess one or two features only. 

In this paper, we take advantage of the separable expression of the beamformer response in the spherical coordinates  
to separate the beamformer design process from the sound field measurement process. We further decompose the beamformer 
design process into several independent parts with each part being responsible for one feature. 
We design a time-domain beamforming method with a frequency-invariant beampattern while focusing on a nearfield point. 
This beamforming method is advantageous over the conventional frequency-domain farfield narrowband beamforming method 
by allowing the enhancement of a broadband nearfield source in real-time.  
The performance of the beamforming method is confirmed by simulation results.

The rest of this paper is organized as follows. We introduce the problem formulation in Sec. II. We derive the beamforming 
method in Sec. III and IV. The effectiveness of the beamforming method is confirmed by simulation in Sec. V, and Sec. VI 
concludes this paper. 

\section{Problem formulation}
\begin{figure}[t]
\centerline{\includegraphics[width=6cm]{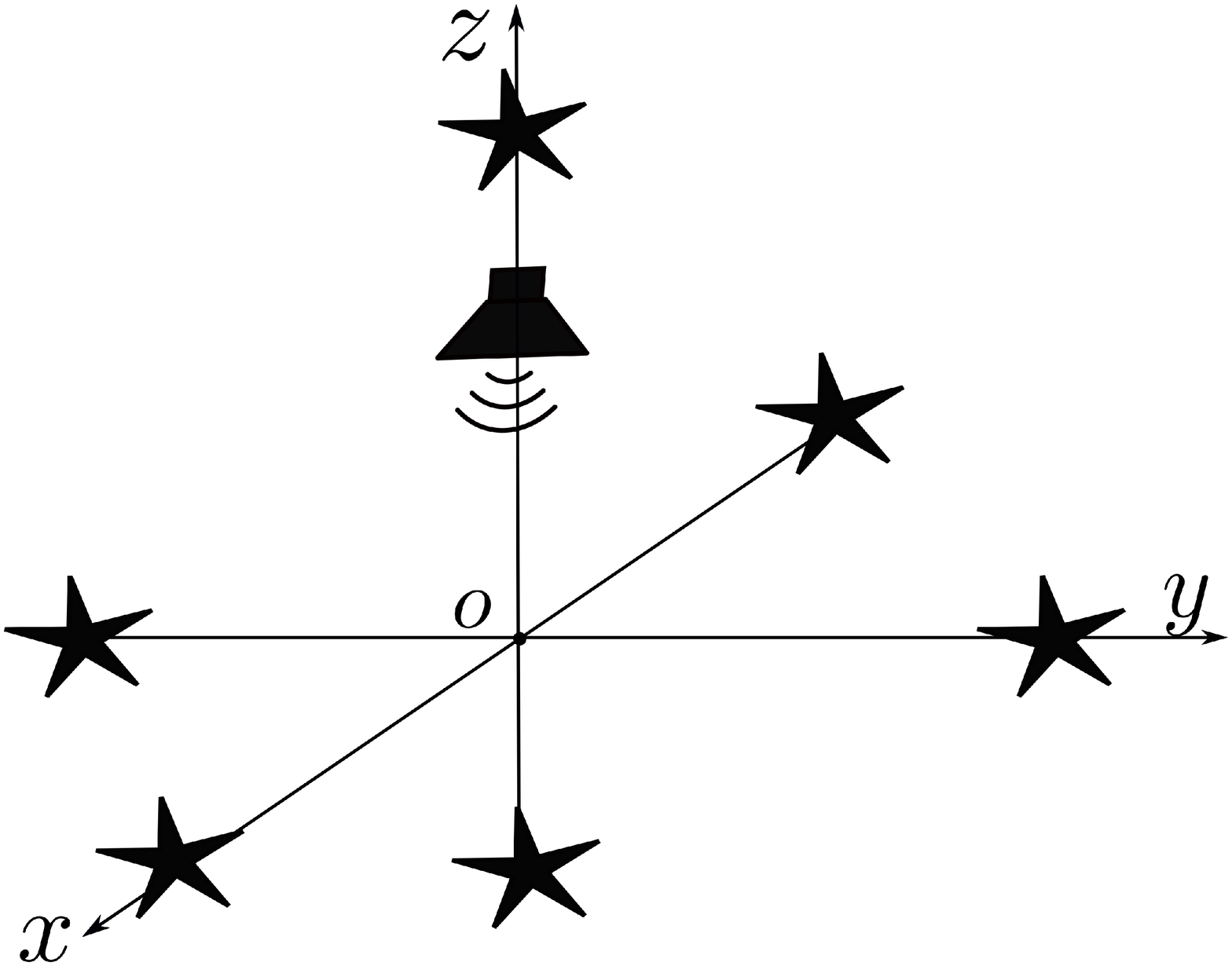}}
\caption{System setup: In the nearfield with respect to the origin $O$, there is a target source (\faVolumeUp) at 
$(r_d,\theta_d,\phi_d)$ and a number of interfering sources ($\star$). We aim to design a time-domain frequency-invariant 
beamformer which focuses on the target nearfield source.} 
\label{fig:nearfield}
\end{figure}

We denote the Cartesian coordinates and the spherical coordinates of a point with respect to an origin $O$ as $(x,y,z)$ 
and $(r,\theta,\phi)$, respectively. 
As shown in Fig.~\ref{fig:nearfield}, in the nearfield with respect to the origin, 
there is a target source (\faVolumeUp) at $(r_d,\theta_d,\phi_d)$ and a number of interfering sources ($\star$). 

In the spherical coordinates, the response of a beamformer to a sound field can be written in the frequency-domain 
as~\cite{Rafaely2015}
\begin{IEEEeqnarray}{rcl}
\label{eq:beam_f}
B(\omega)&=&\sum_{u=0}^{\infty}\sum_{v=-u}^{u}\mathsf{W}_{u,v}(\omega)\mathsf{K}_{u,v}(\omega),
\end{IEEEeqnarray}
where 
\begin{IEEEeqnarray}{rcl}
\mathsf{W}_{u,v}(\omega)=\int_{\Omega}W(\omega,\theta,\phi)Y_{u,v}(\theta,\phi)d\Omega, 
\end{IEEEeqnarray}
\begin{IEEEeqnarray}{rcl}
\mathsf{K}_{u,v}(\omega)=\int_{\Omega}K(\omega,\theta,\phi)Y_{u,v}(\theta,\phi)d\Omega,
\end{IEEEeqnarray}
$\int_{\Omega}(\cdot)Y_{u,v}(\theta,\phi)d\Omega= 
\int_{0}^{2\pi}\int_{0}^{\pi}(\cdot)Y_{u,v}(\theta,\phi)\sin\theta{}d\theta{d}\phi$,
$\omega=2\pi{f}$ is the angular frequency ($f$ is the frequency), 
$Y_{u,v}(\theta,\phi)$ is the spherical harmonic of order $u$ and degree $v$, 
$W(\omega,\theta,\phi)$ is the beamformer sensitivity function, 
$K(\omega,\theta,\phi)$ characterizes the sound field,  
$\mathsf{W}_{u,v}(\omega)$ are beamforming coefficients,
and $\mathsf{K}_{u,v}(\omega)$ are sound field coefficients. 
The beamformer (coefficients) filters the sound field (coefficients) and thus that the beamformer response $B(\omega)$ 
meet some targets such as the directivity index, nearfield focusing, or frequency-invariant \cite{Rafaely2015}. 
Note that we have to truncate the summation over $u$ in \eqref{eq:beam_f} up to some order $N$ 
for practical implementations~\cite{Ward2001-gs,Gumerov2005-up,Kennedy2007-ca}. 

As shown by \eqref{eq:beam_f}, in the spherical coordinates, the beamformer response is the summation of the 
product between of two independent quantities, the sound field (coefficients) and the beamformer 
(coefficients). 

In this work, we aim to design a time-domain frequency-invariant beamforming method, which focus on the 
target nearfield source. 
We first present a method to measure the sound field, and then derive a nearfield frequency-invariant 
beamforming method in the frequency-domain, and at last derive the corresponding time-domain method. 

\section{Sound field measurement}
In this section, we present a method to measure the sound field (coefficients). 
In the frequency-domain, the sound pressure and the radial particle velocity at $(r,\theta,\phi)$ can 
be expressed as~\cite{Williams1999}
\begin{IEEEeqnarray}{rcl}
P(\omega,r,\theta,\phi)
&=&\sum_{u=0}^{\infty}\sum_{v=-u}^{u}\mathsf{P}_{u,v}(\omega,r)Y_{u,v}(\theta,\phi) \nonumber	\\
&=&\sum_{u=0}^{\infty}\sum_{v=-u}^{u}\mathsf{K}_{u,v}(\omega)
{j}_{u}(\omega{}\tau)Y_{u,v}(\theta,\phi),					   
\end{IEEEeqnarray}
and
\begin{IEEEeqnarray}{rcl}
\label{eq:velocity}
V(\omega,r,\theta,\phi)
&=&\frac{i}{\rho{}\omega}
\frac{\partial{ P(\omega,r,\theta,\phi)}}{\partial{r}}							\nonumber\\
&=&
\frac{1}{\rho{c}}
\sum_{u=0}^{\infty}\sum_{v=-u}^{u}
{i}\mathsf{K}_{u,v}(\omega)j_u^{\prime}(\omega\tau)
Y_{u,v}(\theta,\phi)                                                            \nonumber\\
&=&\frac{1}{\rho{}c}\sum_{u=0}^{\infty}\sum_{v=-u}^{u}\mathsf{V}_{u,v}(\omega,r)Y_{u,v}(\theta,\phi),
\end{IEEEeqnarray}
respectively,
where $\tau=r/c$, $c$ is the speed of sound propagation, $i=\sqrt{-1}$ is the unit-imaginary number, $\rho{}$ is the ambient 
density, $\mathsf{P}_{u,v}(\omega,r)$  and $\mathsf{V}_{u,v}(\omega,r)$ are the pressure field coefficients and  velocity 
field coefficients, respectively, 
$j_u(\cdot)$ and $j_u^{\prime}(\cdot)$ are the spherical Bessel function of order $u$ and its derivative with 
respect to the argument, respectively.  

We obtain the radial-dependent pressure field coefficients $\mathsf{P}_{u,v}(\omega,r_s)$ and velocity field coefficients 
$\mathsf{V}_{u,v}(\omega,r_s)$ through 
\begin{IEEEeqnarray}{rcl}
\label{eq:pressure_s}
\mathsf{P}_{u,v}(\omega,r_s)&\approx&\sum_{q=1}^{Q}\gamma_q{}P(\omega,r_s,\theta_q,\phi_q)Y_{u,v}(\theta_q,\phi_q), \\
\label{eq:velocity_s}
\mathsf{V}_{u,v}(\omega,r_s)&\approx&\rho{c}\sum_{q=1}^{Q}\gamma_q{}V(\omega,r_s,\theta_q,\phi_q)Y_{u,v}(\theta_q,\phi_q),
\end{IEEEeqnarray}
where $\{r_s,\theta_q,\phi_q\}_{q=1}^{Q}$ are the spherical coordinates of $Q$ vector sensors~\cite{Vector2003}, which are 
arranged on a sphere of radius $r_s$~\cite{Rafaely2015}, and $\{\gamma_q\}_{q=1}^{Q}$ are the sampling weights. Hereafter 
we use the nearly uniform sampling scheme only, and $\gamma_q=4\pi/Q$ for $q\in[1, Q]$~\cite{Rafaely2015}.

We further obtain the radial-independent sound field coefficients $\mathsf{K}_{u,v}(\omega)$ as 
(please refer \cite{Ma2018} for the details)
\begin{IEEEeqnarray}{rcl}
\label{eq:Kwuv}
\mathsf{K}_{u,v}(\omega)
&=& (i\omega\tau_s)^2\mathsf{V}_{u,v}(\omega,r_s)h_u(\omega\tau_{s})\nonumber \\
&&+ i(\omega\tau_s)^2\mathsf{P}_{u,v}(\omega,r_{s})h^{\prime}_u(\omega\tau_{s}),		
\end{IEEEeqnarray}
where $\tau_s=r_s/c$,  $h_u(\cdot)$ and $h^{\prime}_u(\cdot)$ are the spherical hankel function of order 
$u$ and its derivative with respect to the argument, respectively.

We have the following comments on the sound field measurement process:
\begin{itemize}
\item We use vector sensors for sound field measurement, and this makes \eqref{eq:Kwuv} do not contain the 
division of the spherical Bessel function~\cite{Ma2018}, and facilitates the development of a time-domain 
beamforming method.

\item The vector sensors are not necessary for implementing the beamforming method. 

By exploiting the scattering effect, we can use ordinary microphones mounted on a rigid surface for beamforming. 
The experimental validation of the proposed beamforming method using an Eigenmike, a rigid 
spherical array with 32 microphones, will be reported in a future publication~\cite{MaWASSPA}.
\end{itemize}
\begin{figure}[t]
\centering
\begin{minipage}[b]{0.9\linewidth}
\centerline{\includegraphics[trim={0.5cm 0cm 1cm 0cm},clip,width=9cm]{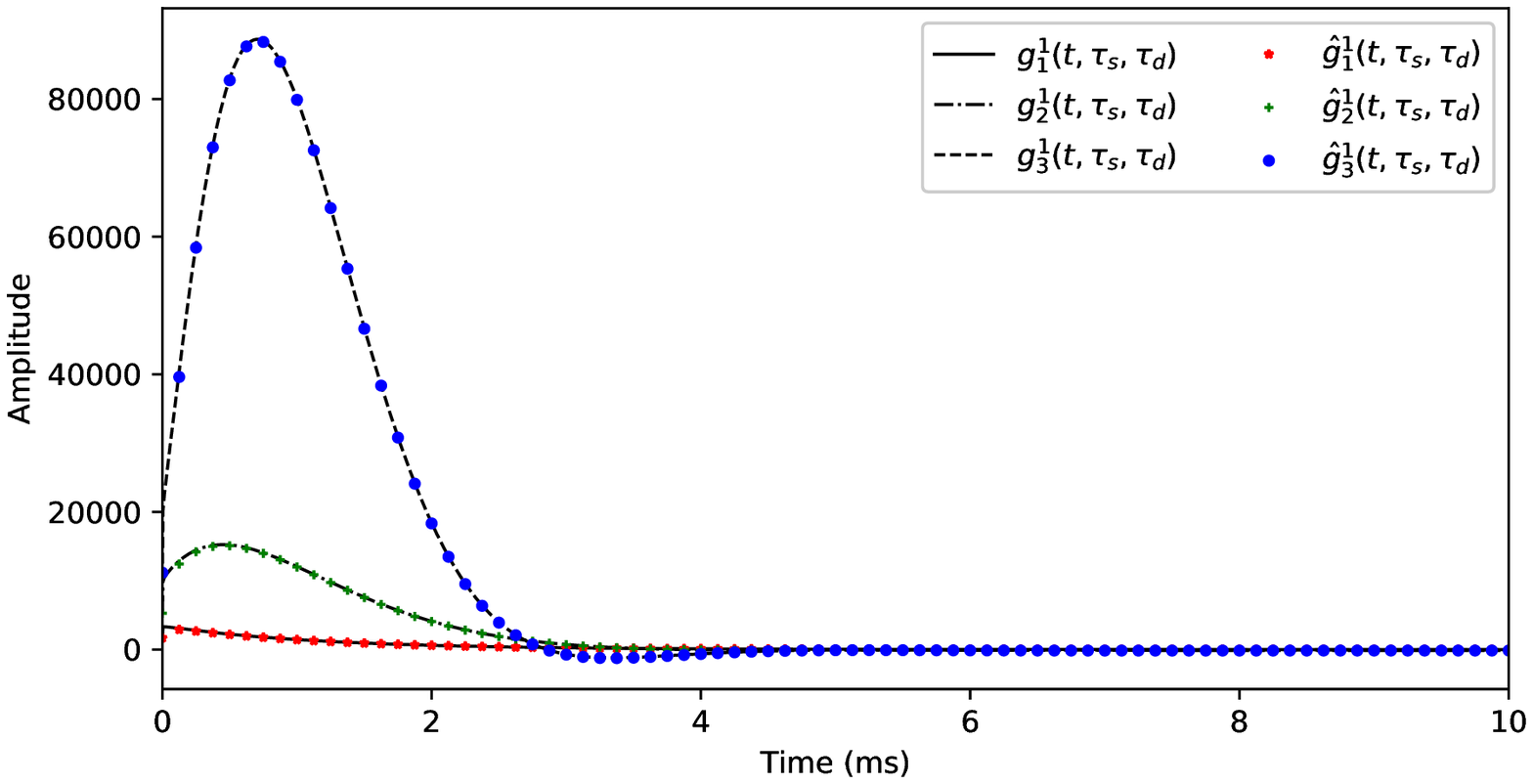}}
\end{minipage}
\caption{The theoretical derived $\mathsf{g}_{u}^1(t,\tau_s,\tau_{\hat{d}})$ and its estimation 
$\hat{\mathsf{g}}_{u}^1(t,\tau_s,\tau_{\hat{d}})$ obtained through direct inverse Fourier transform,
for $u=1,2,3$, $\tau_s=0.23$ ms, and $\tau_{\hat{d}}=1.17$ ms.} \label{fig:gu2t}
\end{figure}

\begin{figure*}[!ht]
\centerline{\includegraphics[width=22cm,angle=90]{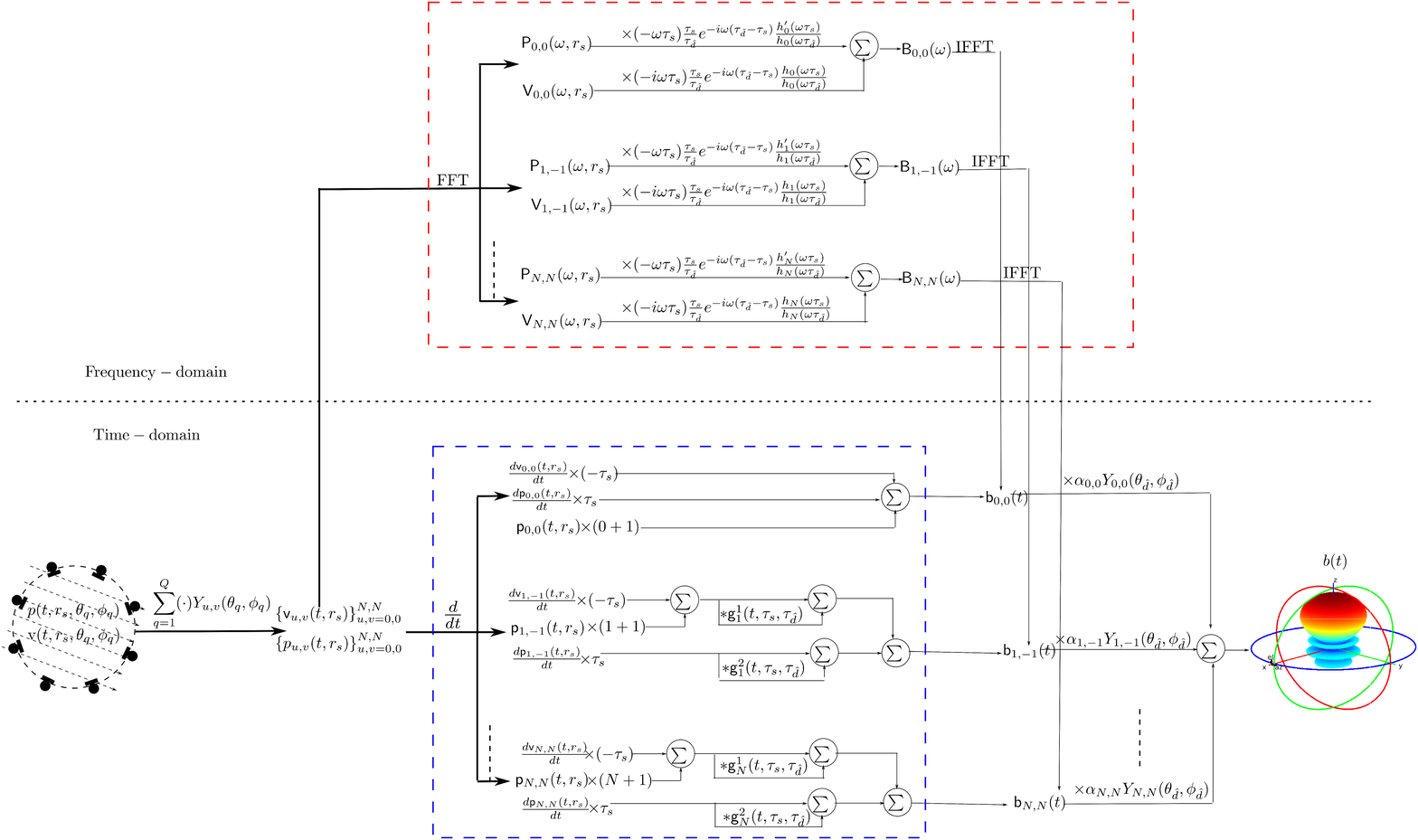}}
\caption{Signal flow of the frequency-domain beamforming method and the time-domain beamforming method. The frequency-domain 
beamforming method uses the FFT and the inverse FFT to transfer the signals between the time-domain and the 
time-frequency-domain. The dashed-line boxes indicate the key differences of the two beamformers.
$\{\cdot\}_{u,v=0,0}^{N,N}$ means $u\in[0,N]$, $v\in[-u,u]$.}
\label{fig:signalflow}
\end{figure*}

\section{Beamforming}
With the sound field measurements available, in this section we develop nearfield frequency-invariant beamforming methods. 

\subsection{\label{sec:freq_beam}Frequency-domain}
In this section, we present a  frequency-domain nearfield frequency-invariant beamforming method.

Denote an estimation of  the target source position as $(r_{\hat{{d}}},\theta_{\hat{{d}}},\phi_{\hat{{d}}})$ 
and define $\tau_{\hat{{d}}}=r_{\hat{{d}}}/c$, 
we design the beamforming coefficients to be 
\begin{IEEEeqnarray}{rcl}
\mathsf{W}_{u,v}(\omega) 
&=&
\alpha_{u,v}
\frac{1}{i\omega\tau_s}                                                                 
\frac{-\tau_s}{\tau_{\hat{{d}}}} e^{-i\omega(\tau_{\hat{d}}-\tau_s)}                     
\nonumber\\
&&
\times 
\frac{1}{h_u(\omega\tau_{\hat{d}})}Y_{u,v}(\theta_{\hat{d}},\phi_{\hat{d}}),         
\label{eq:4dfiltering}
\end{IEEEeqnarray}
where 
\begin{enumerate}
\item $\alpha_{u,v}$ are beampattern coefficients, which determine the beam shape~\cite{Thushara_near}; 
\footnote{Note that for farfield beamforming the beampattern coefficients $\alpha_{u,v}$ 
are the same as the beamforming coefficients $\mathsf{W}_{u,v}(\omega)$.
In this paper of nearfield beamforming, we incorporate the radial-focusing terms $1/h_u(\omega{\tau}_{\hat{d}})$ and 
the spectral filtering term $1/(i\omega\tau_s)$ into the beamforming coefficients $\mathsf{W}_{u,v}(\omega)$.
Thus, we distinguish the beamforming coefficients $\mathsf{W}_{u,v}(\omega)$ from the beampattern coefficients 
$\alpha_{u,v}$. 
}
\item ${1}/({i\omega\tau_s})$ provides spectral filtering, making the beampattern
frequency-invariant~\cite{Dbward}; 
\item $({-\tau_s}/{\tau_{\hat{d}}})e^{-i\omega(\tau_{\hat{d}}-\tau_s)}$ provides gain scale and phase shift, making the 
time-domain spatial functions derived in the next section to be causal~\cite{Yan2011};
\item ${1}/{h_u(k\tau_{\hat{d}})}$ and  $Y_{u,v}(\theta_{\hat{d}},\phi_{\hat{d}})$ provide radial 
filtering~\cite{Thushara_near,rafely-radial} and angular filtering, respectively, making the beamformer to focus on the 
nearfield point $(r_{\hat{d}},\theta_{\hat{d}},\phi_{\hat{d}})$.
\end{enumerate}

Bringing \eqref{eq:4dfiltering} and \eqref{eq:Kwuv} into \eqref{eq:beam_f}, we have the frequency-domain beamformer 
response 
\begin{IEEEeqnarray}{rcl}
\label{eq:beam_f_2}
B(\omega) 
&=&\sum_{u=0}^{\infty}\sum_{v=-u}^{u}
\alpha_{u,v}
\mathsf{B}_{u,v}(\omega)Y_{u,v}(\theta_{\hat{d}},\phi_{\hat{d}} )	
,
\end{IEEEeqnarray}
where 
\begin{IEEEeqnarray}{rcl}
\label{eq:beam_f_uv}
\mathsf{B}_{u,v}(\omega)
&=&
\mathsf{V}_{u,v}(\omega,r_s)          
(-i\omega\tau_s)\frac{\tau_s}{\tau_{\hat{d}}} e^{-i\omega(\tau_{\hat{d}}-\tau_s)} 
\frac{h_u(\omega\tau_{s})}{h_u(\omega\tau_{\hat{d}})}                           
\nonumber\\
&&+
\mathsf{P}_{u,v}(\omega,r_{s}) 
(-\omega\tau_s)
\frac{\tau_s}{\tau_{\hat{d}}} e^{-i\omega(\tau_{\hat{d}}-\tau_s)}               
\frac{h^{\prime}_u(\omega\tau_{s})}{h_u(\omega\tau_{\hat{d}})}. \IEEEeqnarraynumspace
\end{IEEEeqnarray}

We provide additional details for the beamforming coefficient design in the appendix.

\subsection{Time-domain}
The frequency-domain beamforming method can be effectively implemented based on the FFT. 
Nonetheless the FFT unavoidably introduces latency (the frame-delay) to the beamformer output, 
and thus that the beamformer output may not be suitable for further use by time-critical applications, 
such as active noise control~\cite{vary2018,anuphone}.
In this section, we present a time-domain nearfield frequency-invariant beamforming method, which 
processes the signals sample-by-sample and thus can achieve the lowest latency between the sensor 
array input and the beamformer output. 

We write the time-domain beamformer response as 
\begin{IEEEeqnarray}{rcl}
\label{eq:beamer_t}
b(t) &=&\sum_{u=0}^{N}\sum_{v=-u}^{u}
\alpha_{u,v}
\mathsf{b}_{u,v}(t)Y_{u,v}(\theta_{\hat{d}},\phi_{\hat{d}} ),
\end{IEEEeqnarray}
where 
\begin{IEEEeqnarray}{rcl}
\label{eq:beamer_uv_t}
\mathsf{b}_{u,v}(t)
&=& \Big [\frac{d\mathsf{v}_{u,v}(t,r_s) }{dt}  +\frac{d\mathsf{v}_{u,v}(t,r_s) }{dt} \ast 
\mathsf{g}_u^1(t,\tau_s,\tau_{\hat{d}}) \big]\times(-\tau_s)                                          \nonumber\\
&&+
\Big[
\frac{d\mathsf{p}_{u,v}(t,r_s) }{dt}  +\frac{d\mathsf{p}_{u,v}(t,r_s) }{dt} \ast 
\mathsf{g}_u^2(t,\tau_s,\tau_{\hat{d}}) \Big] \times\tau_s                                         \nonumber\\
&&+\Big[\mathsf{p}_{u,v}(t,r_{s}) +\mathsf{p}_{u,v}(t,r_{s}) \ast \mathsf{g}_u^1(t,\tau_s,\tau_{\hat{d}}) 
        \Big ]\times(u+1),                     \nonumber\\
\end{IEEEeqnarray}
$\mathsf{p}_{u,v}(t,r_{s})$ and ${d\mathsf{p}_{u,v}(t,r_s) }/{dt}$ are the time-domain pressure field coefficients and 
their derivation with respect to time $t$, respectively, 
$\mathsf{g}_u^1(t,\tau_s,\tau_{\hat{d}})$ and  $\mathsf{g}_u^2(t,\tau_s,\tau_{\hat{d}})$ are spatial functions, which 
realize the spectral filtering and the radial filtering as time-domain filters. The detailed derivations of above equation 
and the explicit expressions of $\mathsf{g}_u^1(t,\tau_s,\tau_{\hat{d}})$ and  $\mathsf{g}_u^2(t,\tau_s,\tau_{\hat{d}})$ are 
in the appendix.

The time-domain pressure field coefficients $\mathsf{p}_{u,v}(t,r_s)$ are obtained through 
\begin{IEEEeqnarray}{rcl}
\label{eq:puvt}
\mathsf{p}_{u,v}(t,r_s)
&=&\int_{\Omega}p(t,r_s,\theta,\phi)Y_{u,v}(\theta,\phi)d\Omega  \nonumber\\
&\approx&\sum_{q=1}^{Q}\gamma_q{}p(t,r_s,\theta_q,\phi_q)Y_{u,v}(\theta_q,\phi_q)   ,
\end{IEEEeqnarray}
where $p(t,r_s,\theta_q,\phi_q)$ is the time-domain pressure at the sampling point $(r_s,\theta_q,\phi_q)$, 
and the sampling process is similar to \eqref{eq:pressure_s}. 
We further obtain ${d\mathsf{p}_{u,v}(t,r_s) }/{dt}$, the time-derivative of $\mathsf{p}_{u,v}(t,r_s)$, through 
the following approximation 
\begin{IEEEeqnarray}{rcl}
\label{eq:difft}
 \frac{d\mathsf{p}_{u,v}(t,r_s) }{dt} 
 &=&\lim_{\delta_t\to0}\frac{\mathsf{p}_{u,v}(t+\delta_t,r_s)- \mathsf{p}_{u,v}(t-\delta_t,r_s) }{2\delta_t} \nonumber\\
 &\approx&\frac{\mathsf{p}_{u,v}(t+\delta_{f_\mathrm{s}},r_s)- \mathsf{p}_{u,v}(t-\delta_{f_\mathrm{s}},r_s) }
 {2\delta_{f_\mathrm{s}}}, 
\end{IEEEeqnarray}
where $\delta_{f_\mathrm{s}}=1/f_\mathrm{s}$ and $f_\mathrm{s}$ is the sampling frequency. 
To minimize the approximation error of \eqref{eq:difft}, we need to choose an appropriate sampling frequency based on the 
frequency-component of the target source~\cite{Ma2018}.
We can obtain the time-domain velocity field coefficients and the corresponding time-derivative similarly.

We have the following comments on the beamforming methods: 
\begin{enumerate}
\item In \eqref{eq:4dfiltering}, we decompose the beamforming coefficients into several independent parts,
and each part is responsible for one feature of the proposed beamforming method.
\item We can approximate the time-derivative of the particle velocity by the pressure gradient~\cite{Ma2018} as follows  
\begin{IEEEeqnarray}{rcl}
 \frac{d{v_x(t,x,y,z)}}{dt}\approx{}\frac{1}{\rho}\frac{p(t,x+\delta_x,y,z)-p(t,x-\delta_x,y,z)}{2\delta_x}, \nonumber \\
\end{IEEEeqnarray}
where $v_x(t,x,y,z)$ denotes the particle velocity at $(x,y,z)$ along the positive $x$-axis direction, 
$p(t,x+\delta-x,y,z)$ and $p(t,x-\delta_x,y,z)$ are the pressures at  $(x+\delta_x,y,z)$ and $(x-\delta_x,y,z)$, 
respectively, and $\delta_x$ is a small positive value. 

That is, we can use two closely spaced pressure microphones in place of a velocity sensor. 

\item For a rotation symmetric beampattern, the beampattern coefficients $\alpha_{u,v}=\alpha_{u,0}$ for $v\in[-u,u]$.  
This simplifies the computation of the beamformer response~\cite{Rafaely2015}. 
\item Equation \eqref{eq:beamer_uv_t} is derived such that the spatial functions 
$\mathsf{g}_u^1(t,\tau_s,\tau_{\hat{d}})$ 
and $\mathsf{g}_u^2(t,\tau_s,\tau_{\hat{d}})$ do not contain the impulsive function $\delta(t)$ 
\cite{residual}, 
which is difficult to implement as a digital filter. 
\item In Fig.~\ref{fig:gu2t}, we present the theoretical derived $\mathsf{g}_{u}^1(t,\tau_s,\tau_{\hat{d}})$ and its 
estimation $\hat{\mathsf{g}}_{u}^1(t,\tau_s,\tau_{\hat{d}})$ obtained through direct inverse Fourier transform of 
\eqref{eq:GU2W}, for $u=1,2,3$, $\tau_s=0.23$ ms, and $\tau_{\hat{d}}=1.17$ ms. As shown in Fig.~\ref{fig:gu2t}, the 
theoretical derived $\mathsf{g}_{u}^1(t,\tau_s,\tau_{\hat{d}})$ agrees well with its estimation. The theoretical derived 
$\mathsf{g}_{u}^2(t,\tau_s,\tau_{\hat{d}})$ also agrees with its estimation and is not shown for brevity. 

\item As shown by Fig.~\ref{fig:gu2t}, $\mathsf{g}_{u}^1(t,\tau_s,\tau_{\hat{d}})\approx0$ for $t>4$ ms, and thus when we 
implement $\mathsf{g}_{u}^1(t,\tau_s,\tau_{\hat{d}})$ and $\mathsf{g}_{u}^2(t,\tau_s,\tau_{\hat{d}})$ as digital filters we 
can truncate them to be of finite length~\cite{Yan2011}. 

Based on \eqref{eq:difft}, and the truncation of $\mathsf{g}_{u}^1(t,\tau_s,\tau_{\hat{d}})$ and 
$\mathsf{g}_{u}^2(t,\tau_s,\tau_{\hat{d}})$ to a finite length $T_n$, we express \eqref{eq:beamer_uv_t}
using discrete time-index $n$ and $n^{\prime}$ as 
\begin{IEEEeqnarray}{rcl}
\label{eq:discrete}
&&\mathsf{b}_{u,v}(n)
\approx 
(-\tau_s) 
\frac{\mathsf{v}_{u,v}(n+1,r_s)- \mathsf{v}_{u,v}(n-1,r_s) }{2\delta_n}    
\nonumber\\
&&
+(u+1)
\mathsf{p}_{u,v}(n,r_{s})                   \nonumber\\
&&+\tau_s  
\frac{\mathsf{p}_{u,v}(n+1,r_s)- \mathsf{p}_{u,v}(n-1,r_s) }{2\delta_n}    \nonumber\\
&&+
\sum_{n^{\prime}=0}^{T_n}
\Big[(-\tau_s) 
\frac{\mathsf{v}_{u,v}(n+1-n^{\prime},r_s)- \mathsf{v}_{u,v}(n-1-n^{\prime},r_s) }{2\delta_n}    \nonumber\\
&&+(u+1)\mathsf{p}_{u,v}(n-n^{\prime},r_{s}) 
\Big] 
\mathsf{g}_u^1(n^{\prime},\tau_s,\tau_{\hat{d}}) \delta_n                                       \nonumber\\
&&+\tau_s  
\sum_{n^{\prime}=0}^{T_n}
\frac{\mathsf{p}_{u,v}(n+1-n^{\prime},r_s)- \mathsf{p}_{u,v}(n-1-n^{\prime},r_s) }{2\delta_n}    \nonumber\\
&&\quad\times \mathsf{g}_u^2(n^{\prime},\tau_s,\tau_{\hat{d}})       \delta_n,                     
\end{IEEEeqnarray}
where $\delta_n=1/f_\mathrm{s}$. 
Bringing \eqref{eq:discrete} back into \eqref{eq:beamer_t}, we obtain the beamformer response $b(n)$ at each discrete time. 

\item 
To facilitate their implementations, we present the signal flow of the frequency-domain beamforming method and 
the time-domain beamforming method in Fig.~\ref{fig:signalflow}.

\end{enumerate}

\section{Simulation results}
In this section, we design near-field beamformers, demonstrate and compare their performance through simulations.

Here are the basic simulation settings. The speed of sound is $c=343$ m/s and the ambient density of air 
is $\rho=1.225$ kg/m$^3$. The source outputs are unit-variance Gaussian white noises filtered by a 64 
tap Butterworth bandpass filter of frequency range $[f_\mathrm{l}=400, f_\mathrm{h}=4000]$ Hz. 
The sampling frequency is $f_\mathrm{s}=48$ kHz. 
We simulate the impulse response including the particle velocity response between the sources and the 
sensors using the free space Green's function and the Euler's equation~\cite{Williams1999}. We add Gaussian 
white noises to the sensor measurements such that the signal to noise power ratio is 30 dB. 
We normalize the maximum beamformer response to  be unity (or 0 dB).
The simulation results are from an average of 100 independent runs. 
These settings are used in all simulations unless otherwise stated. 


\begin{figure}[t]
\centering
\begin{minipage}[b]{0.8\linewidth}
\centerline{\includegraphics[width=7.5cm]{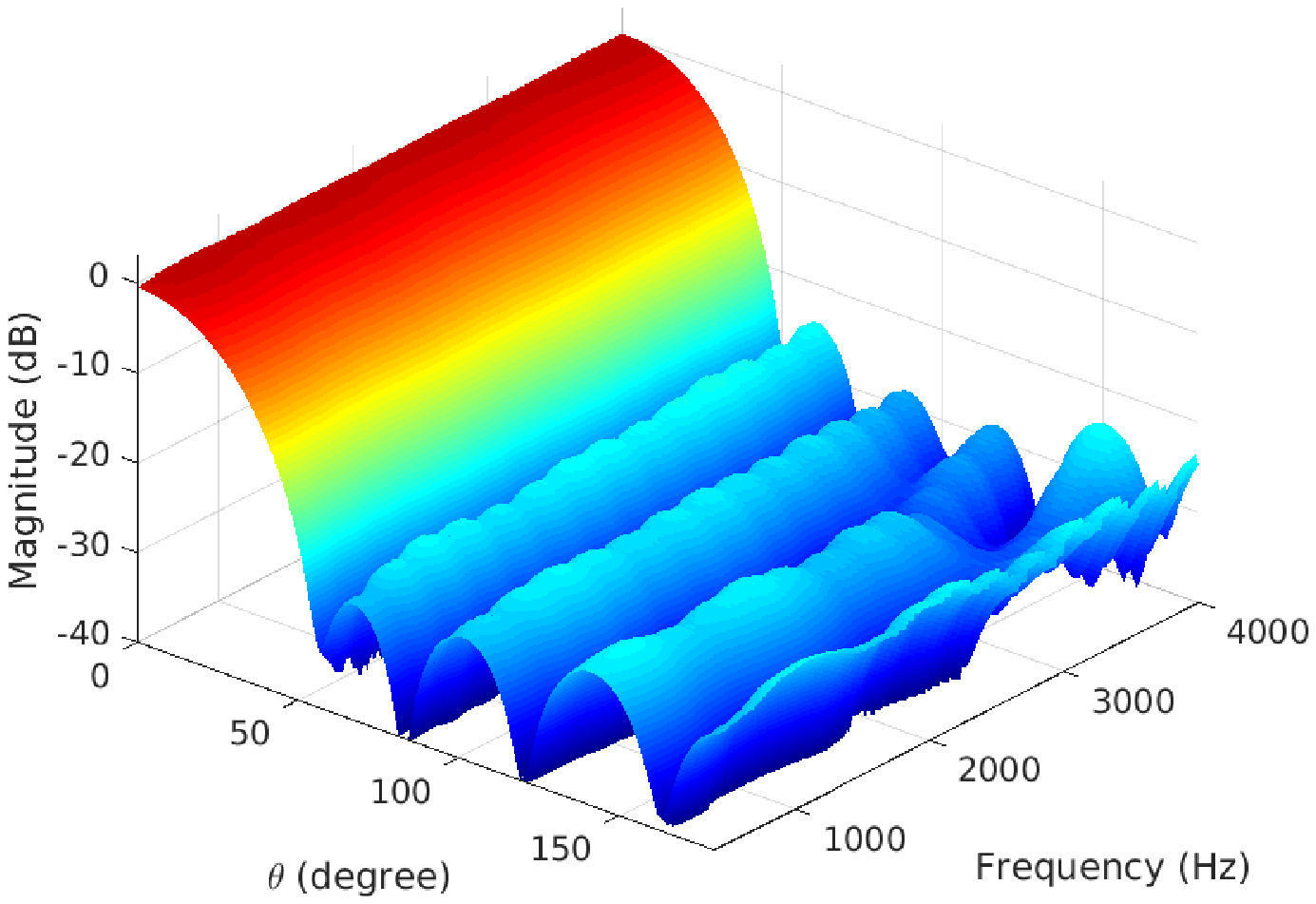}}
\centerline{(a)}
\end{minipage}
\centering
\begin{minipage}[b]{0.8\linewidth}
\centerline{\includegraphics[width=7.5cm]{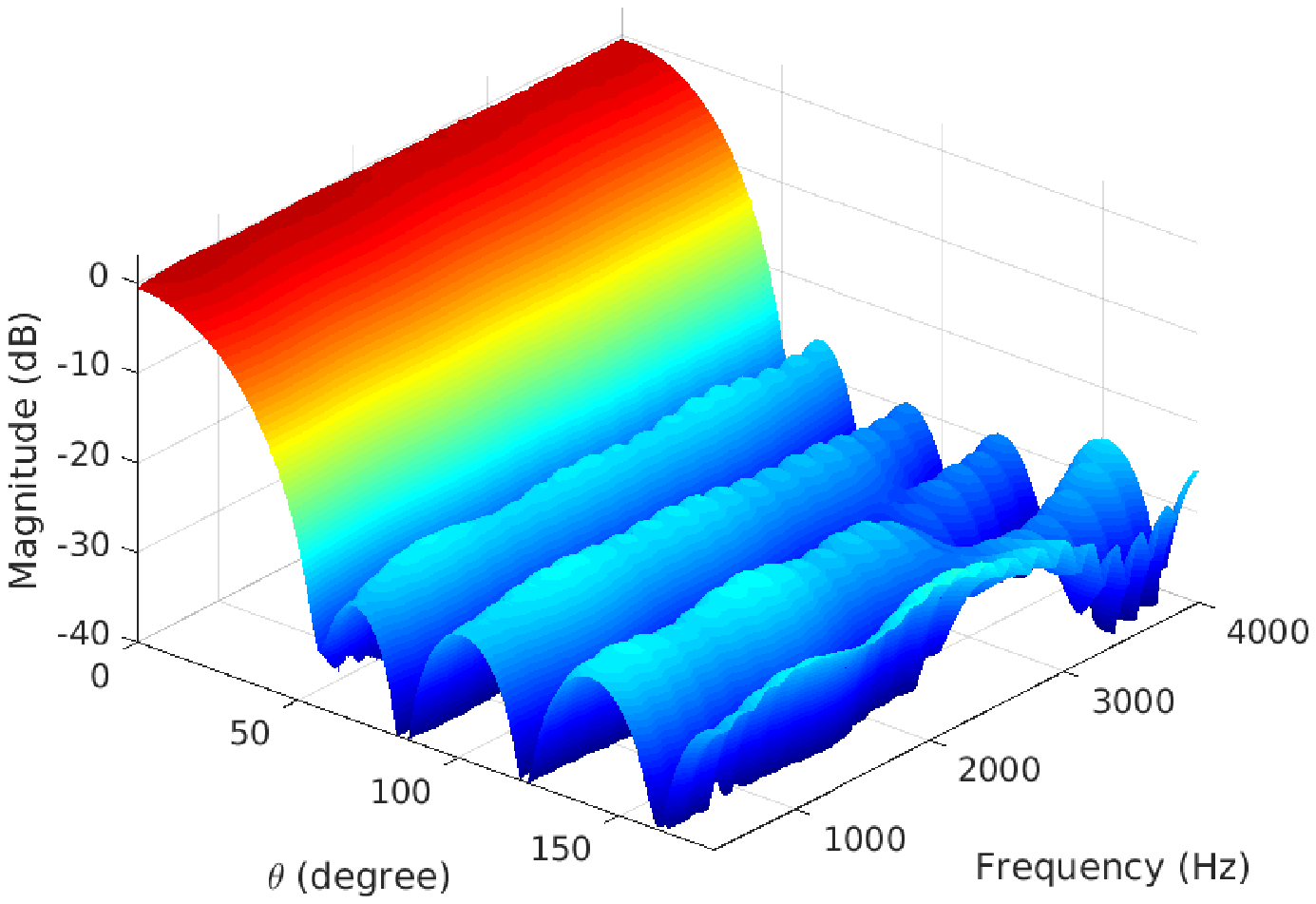}}
\centerline{(b)}
\end{minipage}
\caption{Beamformer response (in dB) to a nearfield point at $(0.4 \; \mathrm{m}, \theta, 0^{\circ})$ as a 
function of frequency and
angle $\theta$: (a) frequency-domain and (b) time-domain.} 
\label{fig:beamft}
\end{figure}

\begin{figure*}[t]
\begin{minipage}[b]{0.5\linewidth}
\centerline{\includegraphics[trim={2cm 3cm 1cm 2cm},clip,width=8cm]{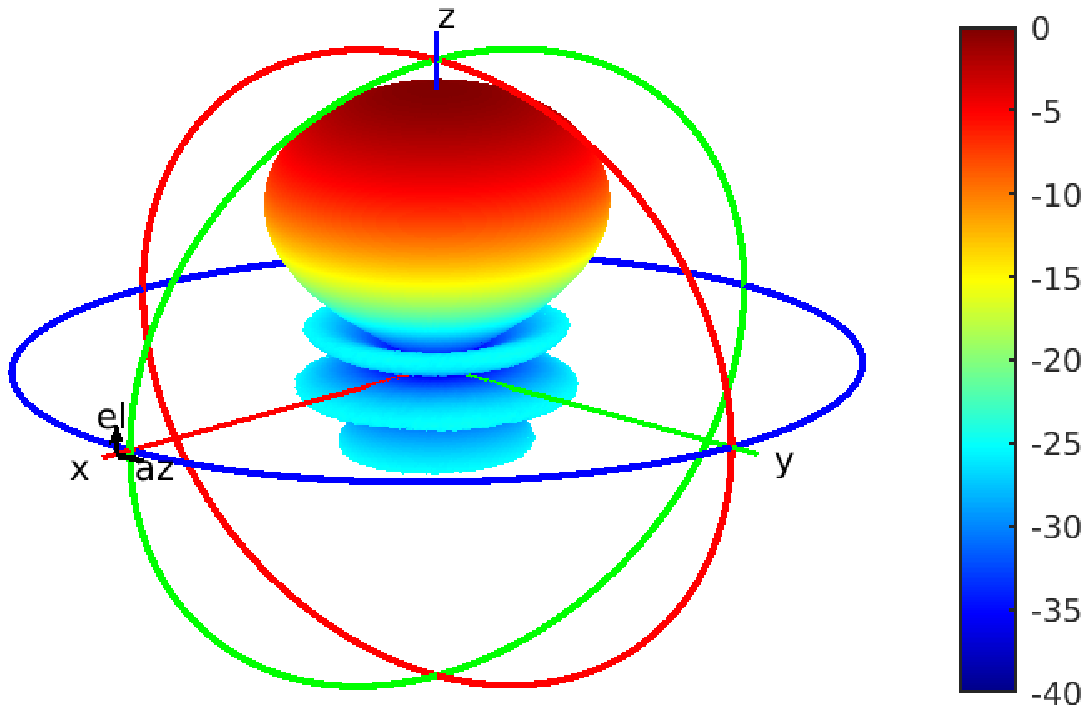}}
\centerline{(a)}
\end{minipage}
\begin{minipage}[b]{0.5\linewidth}
\centerline{\includegraphics[trim={2cm 3cm 1cm 2cm},clip,width=8cm]{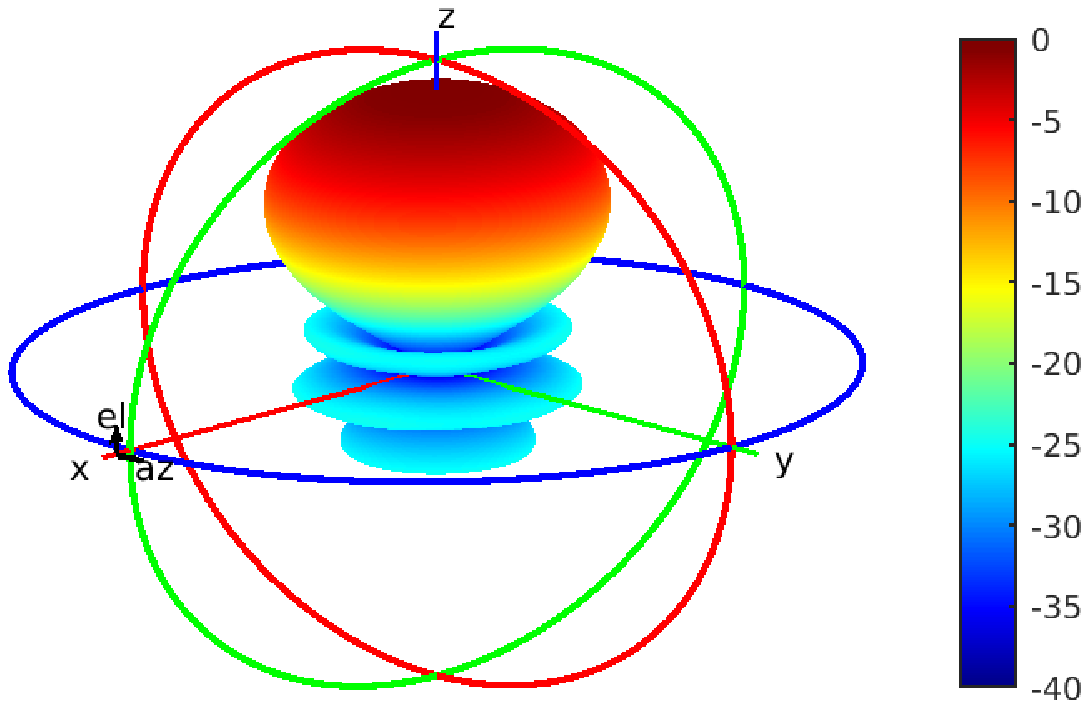}}
\centerline{(b)}
\end{minipage}
\begin{minipage}[b]{0.5\linewidth}
\centerline{\includegraphics[trim={2cm 3cm 1cm 2cm},clip,width=8cm]{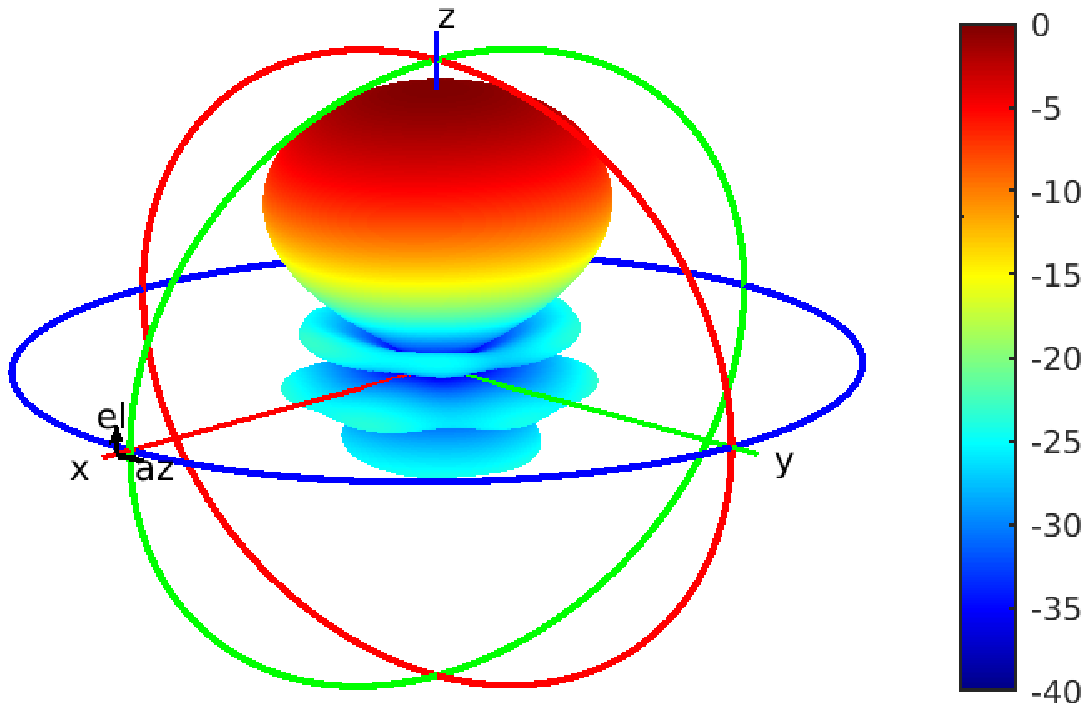}}
\centerline{(c)}
\end{minipage}
\begin{minipage}[b]{0.5\linewidth}
\centerline{\includegraphics[trim={2cm 3cm 1cm 2cm},clip,width=8cm]{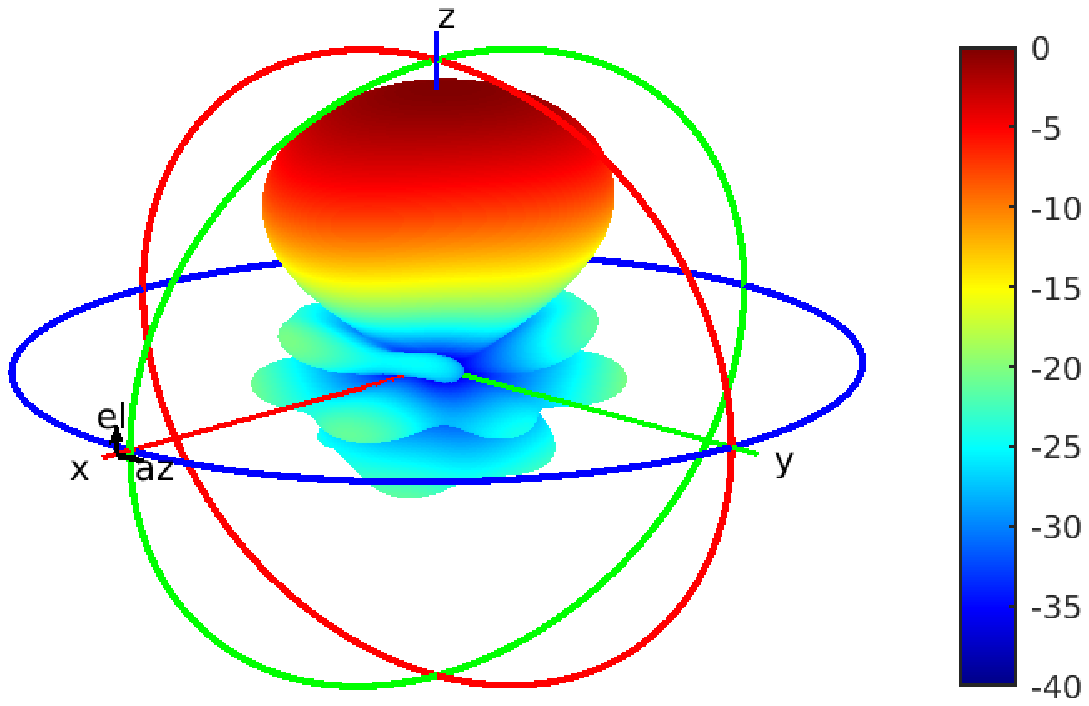}}
\centerline{(d)}
\end{minipage}
\caption{Beamformer response (in dB) to a nearfield point at $(0.4\;\mathrm{m}, \theta, \phi)$  at 
(a) 500, (b) 1500, (c) 2500, and (d) 3500 Hz.}
\label{fig:beam_3d}
\end{figure*}

\subsection{\label{sec:beam_response}Beamformer response}
In this section, we simulate the beamformer response to  nearfield and  farfield points.

Let there be a sphere of radius $r_s=0.08$ m ($\tau_s=r_s/c\approx0.23$ ms) centered at the origin. 
We arrange 36 vector sensors on the sphere according to the nearly uniform sampling scheme 
\cite{Rafaely2015}. 
We design the beampattern coefficients $\alpha_{u,v}$ up to fourth order according to the 
Dolph-Chebyshev beampattern with a -25 dB sidelobe level \cite{rafely-radial,dolph}. 
We design the beamformer to focus at a nearfield point $(r_d,\theta_d,\phi_d)=(0.4\;\mathrm{m}, 0^{\circ}, 
0^{\circ})$ ($\tau_d=r_d/c\approx1.17$ ms).

For the frequency-domain beamforming method, we use FFT  with a block size of 1024 and half overlapping to 
transfer the signals between time-domain and the time-frequency-domain~\cite{Shynk1992-af}.  
To implement the time-domain beamforming method, based on Fig.~\ref{fig:gu2t}, we truncate 
$\mathsf{g}_u^1(t,\tau_s,\tau_{\hat{d}})$ and 
$\mathsf{g}_u^2(t,\tau_s,\tau_{\hat{d}})$ to be $L=0.005*f_\mathrm{s}=240$ taps. 

We present the frequency-domain  beamformer response to a nearfield point at $(0.4\;\mathrm{m}, \theta, 0^{\circ})$ as a 
function of frequency and angle in Fig.~\ref{fig:beamft} (a), which shows  a constant mainlobe width of [-$50^{\circ}$, 
$50^{\circ}$]  over [400, 4000] Hz. 
(The beamformer response is symmetric with respect to $\theta=90^{\circ}$, and thus in the figure the other half is not shown 
for brevity.) The sidelobe level is less than -25 dB between [400, 4000] Hz except that at  $\theta=65^{\circ}, 160^{\circ}$ 
the sidelobe level is about $-24$ dB for frequency close to 4000 Hz. 
We present the time-domain beamformer response to the same near field point in Fig.~\ref{fig:beamft} (b), which is similar to 
the frequency-domain beamformer response in Fig.~\ref{fig:beamft} (a).

We present the beamformer response to a nearfield point at $(0.4\;\mathrm{m}, \theta, \phi)$ at 500, 1500, 2500,
and 3500 Hz in Fig.~\ref{fig:beam_3d} (a), (b), (c), and (d), respectively. 
Figure~\ref{fig:beam_3d} shows that the sidelobe levels at the high frequency of $3500$ Hz can be larger than the 
designed -25 dB. Nonetheless, the overall beamformer responses, especially the mainlobe, are similar across all 
these four frequencies. This simulation result together with Fig.~\ref{fig:beamft} demonstrate that the designed 
beamformer has a nearly invariant beam response (beampattern) over a wide frequency band.

\begin{figure}[t]
\begin{minipage}[b]{0.8\linewidth}
\centerline{\includegraphics[width=8.5cm]{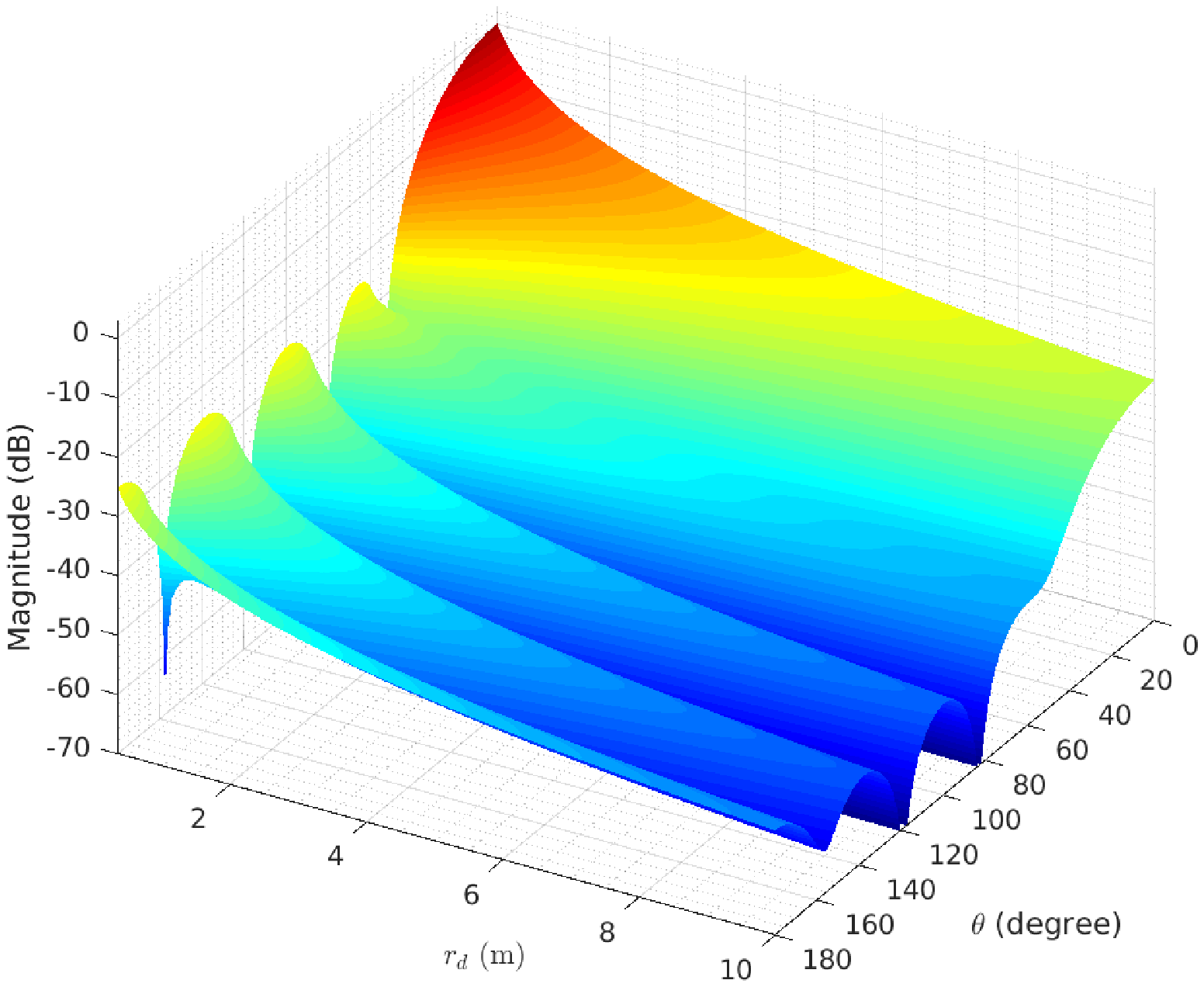}}
\end{minipage}
\caption{Beamformer response (in dB) to a point at $(r_d, \theta, 0^{\circ})$ at $f=1000$ Hz as a function of $r_d$
and angle $\theta$.} 
\label{fig:beam_radial}
\end{figure}

Next we present the response of the designed nearfield beamformer to a point at $(r_d, \theta, 0^{\circ})$ and at $f=1000$ 
Hz in 
Fig.~\ref{fig:beam_radial}. As shown in the figure, unless the point is close to the designed focusing point 
$(0.4\;\mathrm{m}, 0^{\circ}, 0^{\circ})$, the beamformer response will be significantly attenuated. This demonstrates 
the superior far-field and off-axis suppression capability of the nearfield beamformer, which will be useful for speech 
acquisition applications in a room environment. (The beamformer response to a point at $(r_d, \theta, 0^{\circ})$ and 
at other frequency $f\in[400, 4000]$ Hz is similar to Fig.~\ref{fig:beam_radial}, and is not shown for brevity.)

\subsection{Comparison}
In this section, we compare the computation complexity and the latency of the frequency-domain method and the time-domain  
method.

In Fig.~\ref{fig:signalflow}, the dashed-line boxes indicate the key differences between the two methods,
and thus we focus on the computations inside of the boxes, specifically the number of multiplication operations.  
The time-domain beamforming method mainly involves the convolution between the decomposition coefficients
with two filters. We denote the filter length as $L$, then the time-domain beamforming method requires about $2L$ 
real-value multiplication operations to obtain one sample of the coefficient $\mathsf{b}_{u,v}(t)$.   
For the frequency-domain beamforming method, we first transform $\{\mathsf{p}_{u,v}(t,r_s),\mathsf{v}_{u,v}(t,r_s)\}$ 
to be $\{\mathsf{P}_{u,v}(\omega,r_s),\mathsf{V}_{u,v}(\omega,r_s)\}$ using the FFT, and then obtain the coefficients 
$\mathsf{B}_{u,v}(\omega)$ through \eqref{eq:beam_f_uv}, and at last transfer $\mathsf{B}_{u,v}(\omega)$ into 
$\mathsf{b}_{u,v}(t)$. 
Using the radix-2 FFT algorithm with half overlapping and block size $M$~\cite{Shynk1992-af}, it takes 
$[(2\times2+1\times4)M\log_2{M}+2\times{}4M\times(f_\mathrm{h}-f_\mathrm{l})/f_\mathrm{s}]/(M/2)\approx(16\log_2M+4)$
real-value multiplication operations to obtain one sample of the coefficient $\mathsf{b}_{u,v}(t)$.   

\begin{table}[t]
\caption{The computational complexity and the latency of the beamforming methods}
\setlength\tabcolsep{2pt}	
\begin{center}	
\begin{tabular}{|c|c|c|c|c|c|c|c|}	     \hline 	
\shortstack{Time-\\domain}	             & $L=240$ &\shortstack{$L=360$}&\shortstack{$L=480$} & $L=960$ \\	
\hline 
\shortstack{Multiplication\\ operations} & 480    & 720                & 960           & 1920        \\	
\hline 
\shortstack{Latency\\(samples)}          & 0      & 0                & 0  	      &0          \\ 	\hline
\shortstack{Frequency-\\domain}&$M=256$  &\shortstack{$M=512$}&\shortstack{$M=1024$}&\shortstack{$M=2048$}\\\hline 
\shortstack{Multiplication\\ operations} &  132  & 148 & 164 & 180        \\	\hline 
\shortstack{Latency\\(samples)}          &  128  & 256 & 512 & 1024  	 \\	\hline
\end{tabular}
\end{center}
\end{table}

Based on above calculations, we list the computational complexity and the latency of two methods in Table I. 
As shown in Table I, the computational complexity of the time-domain beamforming method is higher than that of 
the frequency-domain beamforming method. 
However, the latency of the time-domain beamforming method is always zero, and this makes the time-domain
beamforming method attractive for time-critical applications, such as active noise 
control~\cite{Ma2018,vary2018,anuphone}.  

\subsection{Application simulation}
In this section, we use the nearfield beamformer for a nearfield source detection application.

\begin{figure}[t]
\centerline{\includegraphics[width=8cm]{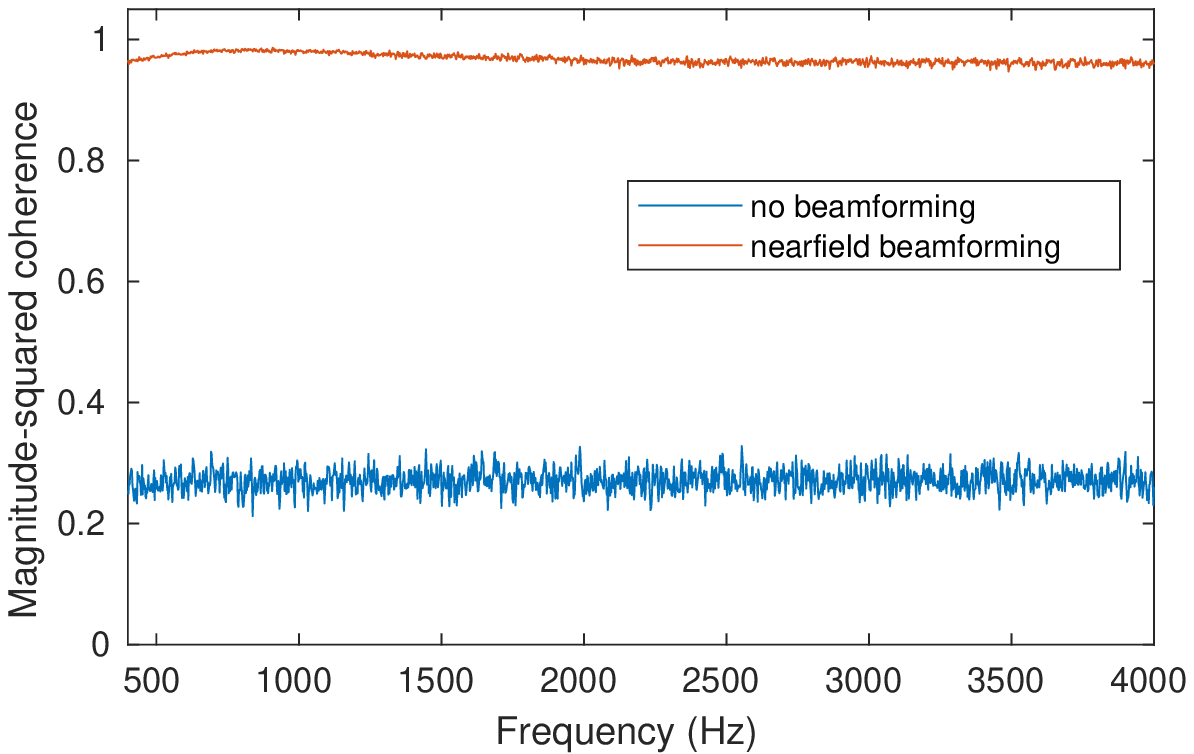}}
\caption{The magnitude-squared coherence between the target source output and the beamformer response.} 
\label{fig:coherence}
\end{figure}

The system setup is as shown in Fig.~\ref{fig:nearfield}, where we have a target point source at $(0.4\;\mathrm{m}, 
0^{\circ}, 
0^{\circ})$ and six interfering point sources at  
$(0.4\;\mathrm{m}, 90^{\circ}, 0^{\circ})$,
$(0.5\;\mathrm{m}, 90^{\circ}, 90^{\circ})$,
$(0.6\;\mathrm{m}, 90^{\circ}, 180^{\circ})$,
$(0.7\;\mathrm{m}, 90^{\circ}, 270^{\circ})$,
$(0.8\;\mathrm{m}, 90^{\circ}, 0^{\circ})$,
and 
$(2.0\;\mathrm{m}, 0^{\circ}, 0^{\circ})$,
respectively. 
Note that one of the interfering source is in the same direction as the target source.
We use the time-domain beamforming method designed in Sec.~\ref{sec:beam_response} to detect the target source output and to 
suppress the interference.   
We denote the target source output and the beamformer response as $p(t)$ and $b(t)$, respectively. 
We perform the simulation over $T=100$ s, and plot the magnitude-square coherence $C_{pb}(f)$ between the 
target source output $p(t)$ and the beamformer response $b(t)$ in Fig.~\ref{fig:coherence}
\begin{IEEEeqnarray}{rcl}
 C_{pb}(f)&=&
 \frac
 {
||P_{pb}(f)||_2^2
 }
 {
 P_{pp}(f)P_{bb}(f)
 },
\end{IEEEeqnarray}
where $||P_{pb}(f)||_2^2$ is the cross power spectral density, 
$ P_{pp}(f)$ and $P_{bb}(f)$ are the auto power spectral densities of $p(t)$ and $b(t)$, respectively~\cite{vary2018}.

As shown in Fig.~\ref{fig:coherence}, the magnitude-squared coherence between the target source output and the 
beamformer response is close to unity over [400, 4000] Hz, which means the beamformer response is highly coherent 
with the target source output, or the beamformer response is mainly contributed by the target source. We also plot 
the coherence between the target source output and the measurement of an omni-directional microphone at the origin, 
which is equivalent to no beamforming process is conducted. As shown in the figure, without beamforming the 
magnitude-squared coherence is lower than 0.3, which means the beamformer response is seriously contaminated by the 
interference. This simulation result demonstrates the nearfield and on-axis suppression capability of the 
beamformer.

\section{Conclusion}
This paper proposed a time-domain beamforming method, whose response is invariant with respect to frequency and 
focuses on a near field points. 
The authors took inspiration from the expression of the beamformer response in the spherical coordinates,
and theoretically derived the beamformer filter (coefficients) with separable terms, which simplify the design
of a beamformer. 
The theoretical derivation can serve as a foundation for future development of robust and data-dependent 
beamforming methods. 

An extension to this paper would be designing similar beamforming methods for two dimensional space. 
In this case, we can use vector sensors arranged on  multiple circles~\cite{Chen2015} or on a line 
\cite{Thusharahigh,Ward1995} for sound field (coefficients) measurement.

\appendices
\section{Frequency-domain}
The beamforming coefficients \eqref{eq:4dfiltering} are designed such that the beamformer response (beampattern) is 
frequency-invariant and focuses on a nearfield point. Suppose that there is a unit-output desired point source at 
$(r_d,\theta_d,\phi_d)$, then we have the corresponding sound field coefficients~\cite{Williams1999}
\begin{IEEEeqnarray}{rcl}
\label{eq:point}
\mathsf{K}_{u,v}(\omega)&=&
(-i\omega/c)h_u(\omega\tau_d)Y_{u,v}(\theta_d,\phi_d).
\end{IEEEeqnarray}
Bringing \eqref{eq:point} and \eqref{eq:4dfiltering} into \eqref{eq:beam_f}, we have the beamformer response to the 
point source as 
\begin{IEEEeqnarray}{rcl}
\label{eq:beamer_p}
B(\omega)
&=&
\frac{-1}{i\omega\tau_s}
\frac{\tau_s}{\tau_{\hat{{d}}}} e^{-i\omega(\tau_{\hat{d}}-\tau_s)}
\frac{-i\omega}{c}      \nonumber\\
&&\times
\sum_{u=0}^{\infty}
\frac{
h_u(\omega\tau_d)
}{h_u(\omega\tau_{\hat{d}})}       
\sum_{v=-u}^{u} 
\alpha_{u,v}
Y_{u,v}(\theta_d,\phi_d) Y_{u,v}(\theta_{\hat{d}},\phi_{\hat{d}})   \nonumber\\
&=& 
\frac{e^{-i\omega(\tau_{{d}}-\tau_s)} }{r_d} 
\delta(\cos\theta_d-\cos\theta_{\hat{d}})\delta(\phi_{d}-\phi_{\hat{d}}), 
\end{IEEEeqnarray}
where $\delta(\cdot)$ is the delta function, and the last line is obtained based on the followings~\cite{Williams1999,hof}
\begin{IEEEeqnarray}{rcl}
\frac{ h_u(\omega\tau_d) }{h_u(\omega\tau_{\hat{d}})} &=& 1,           \\
\sum_{u=0}^{\infty}\sum_{v=-u}^{u} \alpha_{u,v}
Y_{u,v}(\theta_d,\phi_d) Y_{u,v}(\theta_{\hat{d}},\phi_{\hat{d}}) 
&=& 
\delta(\cos\theta_d-\cos\theta_{\hat{d}})   \nonumber\\
&&\times\delta(\phi_{d}-\phi_{\hat{d}}), 
\end{IEEEeqnarray}
given that 
\begin{IEEEeqnarray}{rcl}
(r_d,\theta_d,\phi_d)=(r_{\hat{d}},\theta_{\hat{d}},\phi_{\hat{d}}), \nonumber\\
\alpha_{u,v}=1, \; u\in[0,\infty), v\in[-u,u]. \nonumber
\end{IEEEeqnarray}
That is, the beamforming coefficients $\mathsf{W}_{u,v}(\omega)$ are designed such that the beamformer response to a point 
source at $(r_d,\theta_d,\phi_d)$ reduces to the delta function~\cite{rafely-radial}.

\section{\label{sec:generalization}Time-domain}

We express the frequency-domain expression \eqref{eq:beam_f_uv} as 
\begin{IEEEeqnarray}{rcl}
\label{eq:beamer_f_uv_2}
\mathsf{B}_{u,v}(\omega)
&=&-(i\omega\tau_s)\mathsf{V}_{u,v}(\omega,r_s)
\frac{\tau_s}{\tau_{\hat{d}}} e^{-i\omega(\tau_{\hat{d}}-\tau_s)}
\frac{h_u(\omega\tau_{s})    
}{h_u(\omega\tau_{\hat{d}})} \nonumber\\
&&-\omega\tau_s\mathsf{P}_{u,v}(\omega,r_{s}) \frac{\tau_s}{\tau_{\hat{d}}} e^{-i\omega(\tau_{\hat{d}}-\tau_s)}   
\frac{
h_{u-1}(\omega\tau_s)
}{
h_u(\omega\tau_{\hat{d}})
} \nonumber\\ 
&&+(u+1)\mathsf{P}_{u,v}(\omega,r_{s}) \frac{\tau_s}{\tau_{\hat{d}}} e^{-i\omega(\tau_{\hat{d}}-\tau_s)}   
\frac{
h_{u}(\omega\tau_s)
}{
h_u(\omega\tau_{\hat{d}})
} \nonumber\\ 
&=&-(i\omega\tau_s)\mathsf{V}_{u,v}(\omega,r_s)  
\Big[
\mathsf{G}_u^1(\omega,\tau_s,\tau_{\hat{d}})+1\Big]
                                \nonumber\\
&&+(i\omega\tau_s)\mathsf{P}_{u,v}(\omega,r_{s}) 
\Big[
\mathsf{G}_u^2(\omega,\tau_s,\tau_{\hat{d}})
+1
\Big] \nonumber\\ 
&&+(u+1)\mathsf{P}_{u,v}(\omega,r_{s}) 
\Big[
\mathsf{G}_u^1(\omega,\tau_s,\tau_{\hat{d}})+1\Big],
\end{IEEEeqnarray}
where
\begin{IEEEeqnarray}{rcl}
\mathsf{G}_u^1(\omega,\tau_s,\tau_{\hat{d}})
&=&
\frac{\tau_s}{\tau_{\hat{d}}} e^{-i\omega(\tau_{\hat{d}}-\tau_s)}
\frac{
h_u(\omega\tau_{s})    
}{h_u(\omega\tau_{\hat{d}})}
-1,                                                                             \nonumber\\
&=&
\frac{
\sum_{v=1}^{u}\varphi_v(u)[\frac{1}{(i\omega\tau_s)^{v}}
-
\frac{1}{(i\omega\tau_{\hat{d}})^{v}}]
}{
\sum_{v=0}^{u}\frac{\varphi_v(u)}{(i\omega\tau_{\hat{d}})^{v}}
},
\end{IEEEeqnarray}
and
\begin{IEEEeqnarray}{rcl}
\mathsf{G}_u^2(\omega,\tau_s,\tau_{\hat{d}})
&=&i
\frac{\tau_s}{\tau_{\hat{d}}} e^{-i\omega(\tau_{\hat{d}}-\tau_s)}   
\frac{
h_{u-1}(\omega\tau_s)
}{
h_u(\omega\tau_{\hat{d}})
}-1                                 \nonumber\\
&=&
\frac{
\sum_{v=0}^{u-1}\frac{\varphi_v(u-1)}{(i\omega\tau_s)^{v}}
-
\sum_{v=0}^{u}\frac{\varphi_v(u)}{(i\omega\tau_{\hat{d}})^{v}}
}
{
\sum_{v=0}^{u}\frac{\varphi_v(u)}{(i\omega\tau_{\hat{d}})^{v}}
}.
\end{IEEEeqnarray}
Here, we introduce $\mathsf{G}_u^1(\omega,\tau_s,\tau_{\hat{d}})$ and $\mathsf{G}_u^2(\omega,\tau_s,\tau_{\hat{d}})$ to 
simplify the expressions. They are are obtained based on the following properties of the spherical Hankel 
function~\cite{hof}
\begin{IEEEeqnarray}{rcl}
\label{eq:hankel_exp}
h_{u} (\omega\tau) &=&
{-i^{u}}{e^{-i\omega\tau}}\sum_{v=0}^{u}\frac{\varphi_v(u)}{(i\omega\tau)^{v+1}},   
\end{IEEEeqnarray}
\begin{IEEEeqnarray}{rcl}
\label{eq:hankel_dev}
h_{u}^{\prime} (\omega\tau) 
&=& h_{u-1}(\omega\tau)-\frac{u+1}{\omega\tau} h_{u}(\omega\tau),  \quad   u>0,          
\end{IEEEeqnarray}
\begin{IEEEeqnarray}{rcl}
\label{eq:hankel_derivate}
\frac{h_u(\omega\tau_{s})}{h_u(\omega\tau_{\hat{d}})}&=&\frac{\tau_{\hat{d}}}{\tau_s} e^{-i\omega(\tau_s-\tau_{\hat{d}})} 
, 
\qquad \omega\to\infty	,
\end{IEEEeqnarray}
\begin{IEEEeqnarray}{rcl}
\frac{h_{u-1}(\omega\tau_{s})}{h_u(\omega\tau_{\hat{d}})}&=&\frac{\tau_{\hat{d}}}{\tau_s}(-i)e^{-i\omega(\tau_s-\tau_{
\hat{d}})} 
,\qquad \omega\to\infty.      
\end{IEEEeqnarray}

Based on \eqref{eq:hankel_exp}, we expand $\mathsf{G}_u^1(\omega,\tau_s,\tau_{\hat{d}})$ as 
\begin{IEEEeqnarray}{rcl}
\label{eq:GU2W}
\mathsf{G}_u^1(\omega,\tau_s,\tau_{\hat{d}})
&=&
\frac{
\mathsf{G}_u^{1,1}(\omega,\tau_s,\tau_{\hat{d}})
}{
\mathsf{G}_u^{1,2}(\omega,\tau_s,\tau_{\hat{d}})
},
\end{IEEEeqnarray}
where 
\begin{IEEEeqnarray}{rcl}
\mathsf{G}_u^{1,1}(\omega,\tau_s,\tau_{\hat{d}})&=&
\sum_{v=1}^{u}\varphi_v(u)[\frac{1}{(i\omega\tau_s)^{v}}
- \frac{1}{(i\omega\tau_{\hat{d}})^{v}}],         \\
\mathsf{G}_u^{1,2}(\omega,\tau_s,\tau_{\hat{d}})
&=& \sum_{v=0}^{u}\frac{\varphi_v(u)}{(i\omega\tau_{\hat{d}})^{v}},
\end{IEEEeqnarray}
and 
\begin{IEEEeqnarray}{rcl}
\mathsf{G}_u^{1,2,\prime}(\omega,\tau_s,\tau_{\hat{d}})
&=& -(i\tau_{\hat{d}})\sum_{v=1}^{u}\frac{v\varphi_v(u)}{(i\omega\tau_{\hat{d}})^{v+1}},
\end{IEEEeqnarray}
is the derivative of $\mathsf{G}_u^{1,2}(\omega,\tau_s,\tau_{\hat{d}})$ with respect to $\omega$. 
Based on the Fourier transform property and the residual theorem~\cite{residual},
we have 
\begin{IEEEeqnarray}{rcl}
\label{eq:gu11t}
\mathsf{g}_u^1(t,\tau_s,\tau_{\hat{d}})&=&
\frac{1}{2\pi}\int_{-\infty}^{\infty} \mathsf{G}_u^1(\omega,\tau_s,\tau_{\hat{d}})e^{i\omega{t}}d\omega \nonumber\\
&=&\mathrm{U}(t)i
\sum_{m=1}^{u}
\frac{
\mathsf{G}_u^{1,1}(\omega_{u,m},\tau_s,\tau_{\hat{d}})
}{
\mathsf{G}_u^{1,2,\prime}(\omega_{u,m},\tau_s,\tau_{\hat{d}})
}e^{i\omega_{u,m}t},
\end{IEEEeqnarray}
where where ${\omega_{u,m}}$  is the $m$-th zero of $\mathsf{G}_u^{1,2}(\omega,\tau_s,\tau_{\hat{d}})$, and 
$\mathrm{U}(\cdot)$ is the step function~\cite{nara2017}.

In a similar way, we expand $\mathsf{G}_u^2(\omega,\tau_s,\tau_{\hat{d}})$ as 
\begin{IEEEeqnarray}{rcl}
\mathsf{G}_u^2(\omega,\tau_s,\tau_{\hat{d}})
&=&
\frac{
\mathsf{G}_u^{2,1}(\omega,\tau_s,\tau_{\hat{d}})
}{
\mathsf{G}_u^{1,2}(\omega,\tau_s,\tau_{\hat{d}})
},
\end{IEEEeqnarray}
where 
\begin{IEEEeqnarray}{rcl}
\mathsf{G}_u^{2,1}(\omega,\tau_s,\tau_{\hat{d}})&=&
\sum_{v=1}^{u-1}\frac{\varphi_v(u-1)}{(i\omega\tau_s)^{v}}
-\sum_{v=1}^{u}\frac{\varphi_v(u)}{(i\omega\tau_{\hat{d}})^{v}}, 
\end{IEEEeqnarray}
and obtain 
\begin{IEEEeqnarray}{rcl}
\label{eq:gu21t}
\mathsf{g}_u^2(t,\tau_s,\tau_{\hat{d}})&=&
\frac{1}{2\pi}\int_{-\infty}^{\infty} \mathsf{G}_u^2(\omega,\tau_s,\tau_{\hat{d}})e^{i\omega{t}}d\omega \nonumber\\
&=&\mathrm{U}(t)i
\sum_{m=1}^{u}
\frac{
\mathsf{G}_u^{2,1}(\omega_{u,m},\tau_s,\tau_{\hat{d}})
}{
\mathsf{G}_u^{1,2,\prime}(\omega_{u,m},\tau_s,\tau_{\hat{d}})
}e^{i\omega_{u,m}t}.
\end{IEEEeqnarray}

Based on \eqref{eq:beam_f_2}, \eqref{eq:beamer_f_uv_2}, and the following  
\begin{IEEEeqnarray}{rcl}
\mathsf{p}_{u,v}(t,r_s)&=&\mathcal{F}^{-1}[\mathsf{P}_{u,v}(\omega,r_s) ], 
\end{IEEEeqnarray}
\begin{IEEEeqnarray}{rcl}
\frac{d\mathsf{p}_{u,v}(t,r_s) }{dt} &=&\mathcal{F}^{-1}[(i\omega)\mathsf{P}_{u,v}(\omega,r_s) ],
\end{IEEEeqnarray}
\begin{IEEEeqnarray}{rcl}
\frac{d\mathsf{v}_{u,v}(t,r_s) }{dt} &=&\mathcal{F}^{-1}[(i\omega)\mathsf{V}_{u,v}(\omega,r_s) ], 
\end{IEEEeqnarray}
\begin{IEEEeqnarray}{rcl}
\mathsf{g}_u^1(t,\tau_s,\tau_{\hat{d}})&=&\mathcal{F}^{-1} [\mathsf{G}_u^1(\omega,\tau_s,\tau_{\hat{d}})],
\label{eq:gu1t}
\end{IEEEeqnarray}
\begin{IEEEeqnarray}{rcl}
\mathsf{g}_u^2(t,\tau_s,\tau_{\hat{d}})&=&\mathcal{F}^{-1} [\mathsf{G}_u^2(\omega,\tau_s,\tau_{\hat{d}})].
\label{eq:gu2t}
\end{IEEEeqnarray}
where $\mathcal{F}$ and $\mathcal{F}^{-1}$ stand for the forward and the inverse Fourier transform, 
respectively, we can arrive at \eqref{eq:beamer_t}.

\begin{thebibliography}{1,0}
\def\enquote#1,{``#1,''}
\expandafter\ifx\csname url\endcsname\relax
\def\url#1{\texttt{#1}}\fi
\expandafter\ifx\csname urlprefix\endcsname\relax\def\urlprefix{URL }\fi
\providecommand{\bibinfo}[2]{#2}
\def\plainquote#1{``#1''}
\providecommand{\noopsort}[1]{}
\providecommand{\switchargs}[2]{#2#1}
\providecommand{\dourl}[1]{\href{http://#1}{\nolinkurl{#1}}}
\providecommand{\dodoi}[1]{doi: \href{http://dx.doi.org/#1}{\nolinkurl{#1}}}
\def\eatspace #1{#1}





\bibitem{Bvan1988} 
B. D. V. Veen, and K. M. Buckley, 
`` Beamforming: a versatile approach to spatial filtering," {\em IEEE ASSP Mag.}, vol. 5, no. 2, pp. 4-24, Apr. 1988.

\bibitem{MSES}
S. Gannot,  E. Vincent, S. Markovich-Golan, and A. Ozerov, ``A consolidated perspective on multimicrophone speech enhancement and source separation," 
{\em IEEE/ACM Trans. Audio Speech Lang. Process.}, vol. 25, no. 4, pp. 692-730, Apr. 2017.


\bibitem{Bvan1997}
B. D. Van Veen, W. van Drongelen, M. Yuchtman, and A. Suzuki, "Localization of brain electrical activity via linearly 
constrained minimum variance spatial filtering," {\em IEEE Trans. Biomed. Eng.}, vol. 44, no. 9, pp. 867-880, Sep. 1997.


\bibitem{beamspeech}
E. A. P. Habets and J. Benesty, “A two-stage beamforming approach for noise reduction and dereverberation,” 
{\em IEEE Trans. Audio, Speech Lang. Process.}, vol. 21, no. 5, pp. 945--958, May 2013.

\bibitem{smartspeaker}
X. Xiao, S. Watanabe, H. Erdogan, L. Lu, J. Hershey, M. L. Seltzer et al. "Deep beamforming networks for multi-channel 
speech recognition," {\em 2016 IEEE International Conference on Acoustics Speech and Signal Processing (ICASSP)}, pp. 
5745-5749, Mar. 2016.

\bibitem{DSP}
\bibinfo{author}{A. V. Oppenheim},
\bibinfo{author}{A. S. Willsky},
and \bibinfo{author}{S. H. Nawab},
\emph{\bibinfo{title}{ Signals and  systems 2nd ed}} (\bibinfo{publisher}
{Prentice Hall}, \bibinfo{address}{New Jersey}, \bibinfo{year}{1997}),
Chap.~\bibinfo{chapter}{3-5}, pp.~\bibinfo{pages}{177--400}.
	
\bibitem{Yan2011}
S. Yan, H. Sun, X. Ma, U. P. Svensson, and C. Hou, “Time-Domain Implementation of Broadband Beamformer in Spherical 
Harmonics Domain,” 
{\em IEEE Trans. Audio Speech Lang. Process.}, vol. 19, no. 5, pp. 1221–1230, Jul. 2011.

\bibitem{Thushara_near}
T. D. Abhayapala, R. A. Kennedy, and R. C. Williamson, “Nearfield broadband array design using a 
radially invariant modal expansion,” 
{\em J. Acoust. Soc. Am.}, vol. 107, no. 1, pp. 392–403, 2000. 

\bibitem{coupling}
L. C. Parra, “Steerable frequency-invariant beamforming for arbitrary arrays,” 
{\em J. Acoust. Soc. Am.}, vol. 119, no. 6, pp. 
3839–3847, 2006.

\bibitem{Dbward}
D. B. Ward, R. A. Kennedy, and R. C. Williamson, ``Theory and design of broadband sensor arrays with 
frequency 
invariant far-field beam patterns,” 
{\em J. Acoust. Soc. Am.}, vol. 97, no. 2, pp. 1023–1034, Feb. 1995.

\bibitem{Liu}
W. Liu S. Weiss J. G. McWhirter and I. K. Proudler, "Frequency invariant beamforming for 
two-dimensional and 
three-dimensional arrays," {\em Signal Process}. vol. 87, pp. 2535-2543, Nov. 2007.

\bibitem{Huang}
G. Huang, J. Chen, and J. Benesty, “Insights into frequency-invariant beamforming with concentric 
circular microphone 
arrays,” {\em IEEE/ACM Trans. Audio Speech Lang. Process.}, vol. 26, no. 12, pp. 2305–2318, Dec. 
2018. 

\bibitem{Rafaely2015}	
B. Rafaely, {\em Fundamentals of Spherical Array Processing}, (New York NY USA:Springer, 2015), vol. 8.

\bibitem{rb1}
R. G. Lorenz and S. P. Boyd, “Robust minimum variance beamforming,” 
{\em IEEE Trans. Signal Process.}, vol. 53, no. 5, pp. 1684–1696, May 2005.

\bibitem{rb2}
S. E. Nai, W. Ser, Z. L. Yu, and H. Chen, “Iterative robust minimum variance beamforming,” 
{\em IEEE Trans. Signal Process.}, 
vol. 59, no. 4, pp. 1601–1611, Apr. 2011.

	\bibitem{Ward2001-gs}
	\bibinfo{author}{D.~B. Ward}
	and \bibinfo{author}{T.~D. Abhayapala},
	\enquote{\bibinfo{title}{{Reproduction of a plane-wave sound field using
				an array of	loudspeakers}}},
	{\bibinfo{journal}{IEEE Trans. Speech Audio Process.}},
	vol. 9, no. 66, pp.~\bibinfo{pages}{697--707}, \bibinfo{year}{2001}.
	
	\bibitem{Gumerov2005-up}
	\bibinfo{author}{N. A. Gumerov} and 
	\bibinfo{author}{R. Duraiswami},
	\emph{\bibinfo{title}{Fast multipole methods for the Helmholtz equation in three dimensions}} 
	(\bibinfo{publisher}{Elsevier},  \bibinfo{year}{2005}).
	
	\bibitem{Kennedy2007-ca}
	\bibinfo{author}{R. A. Kennedy}, and 
	\bibinfo{author}{P. Sadeghi}, and 
	\bibinfo{author}{T. D. Abhayapala}, and
	\bibinfo{author}{H. M. Jones},
	\enquote{\bibinfo{title}{Intrinsic limits of dimensionality and richness in random
			multipath fields}}, 
	\bibinfo{journal}{IEEE Trans. Signal Process.}, vol. 55, no. 6,~pp.~\bibinfo{pages}{2542--2556}, 
	\bibinfo{year}{2007}.
	
\bibitem{Williams1999}
E. G. Williams, {\em Fourier Acoustics: Sound Radiation and Nearfield Acoustical Holography}, (Academic Press, 1999).



\bibitem{Vector2003}
H. E. De Bree, “The Microflown: An Acoustic Particle Velocity Sensor,” 
{\em Acoust. Aust.}, vol. 31, 
no. 3, pp. 91–94, Dec. 2003.

\bibitem{Ma2018}
F. Ma, W. Zhang, and T. D. Abhayapala, “Real-time separation of non-stationary sound fields on spheres,” 
{\em J. Acoust. Soc. Am.}, vol. 146, no. 1, p. 11, Jul. 2019.

\bibitem{MaWASSPA}
F. Ma, T. D. Abhayapala, and Prasanga N. Samarasinghe, 
“Time-domain nearfield frequency-invariant beamforming with a rigid spherical array,” 
{\em 2021 IEEE Workshop on Applications of Signal Processing to Audio and Acoustics (WASPAA)}, to be submitted.



\bibitem{rafely-radial}	
E. Fisher and B. Rafaely, “Near-Field Spherical Microphone Array Processing With Radial Filtering,” 
{\em IEEE Trans. Audio Speech Lang. Process.}, vol. 19, no. 2, pp. 256–265, Feb. 2011.	

	\bibitem{vary2018}
	\bibinfo{author}{S.~M. Kuo}, and
	\bibinfo{author}{D.~R. Morgan}, 
	\emph{\bibinfo{title}{Active noise control systems: algorithms and {DSP} implementations}} 
	(\bibinfo{publisher}{John Wiley \& Sons, Inc.}, \bibinfo{address}{New York}, \bibinfo{year}{1995}).
	
	\bibitem{anuphone} 
	\bibinfo{author}{S. Liebich},
	\bibinfo{author}{J. Fabry},
	\bibinfo{author}{P. {Jax}}, and 
	\bibinfo{author}{P. {Vary}}, 
	\enquote{\bibinfo{title}{{Signal Processing Challenges for Active Noise Cancellation Headphones}}}, 
	{\bibinfo{journal}{Speech Communication; 13th ITG-Symposium}}, pp.~\bibinfo{pages}{1--5}, 
	\bibinfo{year}{2018}.


\bibitem{residual}
T. B. Hansen, ``Spherical expansions of time‐domain acoustic fields: Application to 
near‐field scanning,'' {\em J. Acoust. Soc. Am.},
vol. {98}, no. {2}, pp. {1204-1215}, {1995}.



\bibitem{dolph}	
\bibinfo{author}{A. Koretz},
and 
\bibinfo{B. Rafaely}, 
\enquote{\bibinfo{title}{Dolph-Chebyshev beampattern design for spherical arrays}},
	{\bibinfo{journal}{IEEE Trans. Signal Process.}},
	vol. {57}, no. 6,~pp.~\bibinfo{pages}{2417--2420}, \bibinfo{year}{2009}.
	

	\bibitem{Shynk1992-af}
	\bibinfo{author}{Shynk, John J},
	\enquote{\bibinfo{title}{{Frequency-domain and multirate adaptive filtering}}},
	{\bibinfo{journal}{IEEE Signal Process. Mag.}},
	vol. {9} no. 1,~pp.~\bibinfo{pages}{14--37}, \bibinfo{year}{1992}.
	




\bibitem{Chen2015}
H. Chen, T. D. Abhayapala, and W. Zhang, "Theory and design of compact hybrid microphone arrays 
on two-dimensional planes for three-dimensional soundfield analysis," 
{\em J. Acoust. Soc. Am.}, vol. 138, 
no. 5, pp. 3081--3092, Nov. 2015.

\bibitem{Thusharahigh}
T. D. Abhayapala and A. Gupta, “Higher order differential-integral microphone arrays,” 
{\em J. Acoust. Soc. Am.}, vol. 127, no. 5, pp. 227–233, May 2010.

\bibitem{Ward1995}
D. B. Ward, R. A. Kennedy, and R. C. Williamson, “Theory and design of broadband sensor arrays with frequency 
invariant far-field beam patterns,” {\em J. Acoust. Soc. Am.}, vol. 97, no. 2, pp. 1023–1034, Feb. 1995.


\bibitem{hof} 
NIST Digital Library of Mathematical Functions. http://dlmf.nist.gov/, Release 1.0.28 of 2020-09-15. F. W. J. 
Olver, A. B. Olde Daalhuis, D. W. Lozier, B. I. Schneider, R. F. Boisvert, C. W. Clark, B. R. Miller, B. V. Saunders, H. 
S. Cohl, and M. A. McClain, eds.

	
\bibitem{nara2017}	
\bibinfo{author}{N. Hahn},
\bibinfo{author}{F. Winter},
and 
\bibinfo{Spors S.}, 
\enquote{\bibinfo{title}{Synthesis of a spatially band-limited plane wave in the time-domain using wave 
		field synthesis}},
{\bibinfo{journal}{25th Eur. Signal Process. Conf. EUSIPCO 2017}},~pp.~\bibinfo{pages}{673--677}, \bibinfo{year}{2017}.


\end{thebibliography}
\end{document}